\documentclass[preprintnumbers,nofootinbib,showkeys,showpacs,amsmath,amssymb]{revtex4}
\usepackage{amsmath,amssymb,graphics,epsfig,subfigure}
\usepackage{color}
\usepackage{multirow}
\usepackage{booktabs}
\usepackage{changes}
\usepackage{footnote}

\newcommand{\tcb}{\textcolor{blue}}

\usepackage{hyperref}
\hypersetup{colorlinks=true,linkcolor=blue,citecolor=magenta}

\DeclareMathOperator{\sgn}{sgn}

\begin{document}
\renewcommand{\baselinestretch}{1.15}

\title{Topology of light rings for extremal and non-extremal Kerr-Newman Taub-NUT black holes without $\mathbb{Z}_2$ symmetry}

\preprint{}

\author{Shan-Ping Wu, Shao-Wen Wei \footnote{Corresponding author. E-mail: weishw@lzu.edu.cn}}

\affiliation{$^{1}$Lanzhou Center for Theoretical Physics, Key Laboratory of Theoretical Physics of Gansu Province, and Key Laboratory of Quantum Theory and Applications of MoE, Lanzhou University, Lanzhou, Gansu 730000, China,\\
 $^{2}$Institute of Theoretical Physics $\&$ Research Center of Gravitation, Lanzhou University, Lanzhou 730000, People's Republic of China}

%\pacs{bla}

%\date{\today}

\begin{abstract}
Understanding the light ring, one kind fundamental orbit, shall provide us with novel insight into the astronomical phenomena, such as the ringdown of binary merger and shadow of black holes. Recently, topological approach has preliminarily demonstrated its potential advantages on the properties of the light rings. However, for the black holes without $\mathbb{Z}_2$ symmetry and extremal spinning black holes are remained to be tested. In this paper, we aim at these two issues. Due to the NUT charge, the Kerr-Newman Taub-NUT solution has no $\mathbb{Z}_2$ symmetry. By constructing the corresponding topology for the non-extremal spinning black holes, we find the topological number keeps unchanged. This indicates that $\mathbb{Z}_2$ symmetry has no influence on the topological number, while it indeed affects the locations of the light rings and deviates them off the equatorial plane. For the extremal spinning black holes, we find its topology is critically dependent of the leading term of the vector's radial component at the zero point of its angular component on the black hole horizon. The findings state that there exists a topological phase transition, where the topological number changes, for the prograde light rings. While no phase transition occurs for the retrograde light rings. Our study uncovers some universal topological properties for the extremal and non-extremal spinning black holes with or without $\mathbb{Z}_2$ symmetry. It also has enlightening significance on understanding the light rings in a more general black hole background.
\end{abstract}


\maketitle
%\newpage
\tableofcontents

\section{Introduction}\label{intro}

The characteristic orbits around the ultra-compact objects and black holes are very important on understanding the nature of strong gravity and quantum gravity. Accompanied by the recent observations of the gravitational waves~\cite{abbott2016observation,abbott2017gw170817,abbott2020gw190521,PhysRevX.9.031040} and black hole images~\cite{akiyama2019firstshadow,akiyama2019firstIm}, such issue is further highlighted. In particular, the light rings of axi-symmetric black hole (or photon sphere in spherically symmetric spacetime) are closely related to the ring down stage of the black hole mergers~\cite{cardoso2009geodesic,berti2014black} and shadows~\cite{Chandrasekhar}.

Similar to the event horizon, the light rings also seem to be intrinsic circular orbits for the photons around the ultra-compact objects or black holes, which is because that the light rings are only dependent of the black hole parameters, while independent of the energy and angular momentum of the photons. For the Schwarzschild black hole, the radius of the photon sphere is 3$M$ with $M$ being the black hole mass, which leads to a shadow with size $3\sqrt{3}M$. The photon sphere shall be broken into two light rings by the black hole spin. If the direction of the orbit angular momentum and black hole spin are the same, we refer them to the prograde light rings which has a smaller radius than that of the radius of the Schwarzschild black hole. While if they are opposite, we call them retrograde light rings with a larger radius. In particular, for different black holes, the sizes of the light rings or photon spheres behave quite differently, which shall imprint potential observed phenomena in black hole mergers or shadows.

Recently, a novel study, topology approach, was first introduced on exploring the ultra-compact object light rings by Cunha, Berti, and Herdeiro~\cite{Cunha:2017qtt}. Soon, this study was generalized to the black hole backgrounds~\cite{cunha2020stationary}. Instead focusing on the size of the light rings, they proposed a topological charge corresponding them. The results imply that, for a four dimensional non-extremal axi-symmetric, asymptotically flat black hole, there exists, at least, one unstable light ring outside the horizon and ergo-surface for each rotation sense. Or, we always have one more unstable light ring than the stable one. Other work on the light ring number can be found in Refs.~\cite{Hod2018plb,Hod2018epjc,Guo,Ghosh,Junior,Sanchis,Hod:2022EBH_LR}. For the Schwarzschild or Kerr black holes, this result is straightforward because there exist only one photon sphere or light ring (for each sense). Nevertheless, we in Ref.~\cite{Wei:TC_BH_PhotonSpheres} showed that when more than one photon spheres are presented, the topological number for the dyonic black hole is -1, indicating that the unstable light ring must exist and its number is one more than the stable one. Taking specific example, we exactly confirmed this result. Although, in the radial direction, the light rings could be stable or unstable, they are all stable in angular direction~\cite{Guo}. For the Schwarzschild-Melvin solution, it was found that there exists a topological phase transition at certain value of the dilatonic coupling due to the modification of the asymptotic behavior at radial infinity~\cite{Junior}. Other related work can also be found in Refs.~\cite{Ghosh,Bargueno:2022EBL_S_LR}. Such topological approach was also generalized to the study of the time-like circular orbits~\cite{Wei:2022mzv,Ye:2023gmk}, black hole thermodynamics~\cite{Wei:2021vdx,Yerra:2022alz,Wu:2022whe,Fan:2022Topologicalphasetransition,Bai:2022Topolthermodynamics,Yerra:2022TopologHPtransition} and so on.

Among the study of the light rings, these black hole solutions considered all possess a $\mathbb{Z}_2$ symmetry, i.e., $\theta \rightarrow \pi - \theta$. This leads to that all the light rings locate on the equatorial plane. One may wonder whether these topological results hold if the $\mathbb{Z}_2$ symmetry of the concerned solutions is broken. Another issue worths to explore is for the extremal black holes. Although it is extensively known that the topological number for the retrograde light rings always holds, the results for the prograde light rings are remained to be tested. These two issues are the main purpose of this paper.

Interestingly, the Taub-NUT black hole solutions without $\mathbb{Z}_2$ symmetry provide a good test to our above concerns. The spinning Kerr-Newman Taub-NUT black holes are characterized by four quantities, the mass $m$, electric charge $q$, spin $a$, and NUT charge $n$\tcb{~\cite{taub1951empty,newman1963empty}}. For non-zero NUT charge, it was found that the spacetime is not globally asymptotically flat for its string singularity~\cite{Misner:1963fr,manko2005physical}, meanwhile, the $\mathbb{Z}_2$ symmetry is also broken. In Refs.~\cite{Wei:Strongfield_GL_KTNspacetime,Pradhan:2014zia,cebeci2016motion,Vandeev:2022Geodesicdeviation}, the influences of the NUT charge on the motion of particles were investigated. The size and distortion of the shadows were also closely related with the NUT charge. Hence, the light rings as one kind characteristic orbits must be influenced. As a result, in this paper, we wonder whether the topology of light rings heavily rely on the NUT charge. Moreover, for the extremal Kerr-Newman Taub-NUT black holes, the topological properties shall also be examined in details.

The organization of this paper is as followings. In Sec.~\ref{spacetimes}, we briefly introduce the solution of the Kerr-Newman Taub-NUT black hole. In Sec.~\ref{TCforBH}, we study the equation of motion for a photon in a non-extremal black hole background, and then the total topological number defining as the sum of winding numbers for zero points are obtained. It was shown that the location of the light rings is shifted off the equatorial plane by the NUT charge. The topological number still remains unchanged. The location of light ring in the $r$-$\theta$ plane is solved in Sec.~\ref{Solution} for both the prograde and retrograde cases. Moreover, for the extremal Kerr-Newman Taub-NUT black holes, we explore their light rings in Sec.~\ref{Extreme BH}. By expanding the constructed vector near the degenerate horizon, we find that the topological number closely depends on the dominate term of the vector's radial component at the zero point of the angular component. For the retrograde case, there is at least one unstable light rings. While for the prograde case, there may be one or no light ring, indicating a topological phase transition. Finally, we summarize and discuss our results in Sec.~\ref{conclusion}.

\section{ Kerr-Newman Taub-NUT spacetime }\label{spacetimes}

Let us start with the black hole solution. In Boyer-Lindquist coordinate, the line element of the Kerr-Newman Taub-NUT black hole can be written as~\cite{Griffiths:2005familyofSolutions,Pradhan:2014zia,cebeci2016motion}
\begin{equation}
	ds^2 = - \frac{ \Delta} {\Sigma} (dt - {\chi}d {\phi})^2
	+ \frac{ \sin^2{\theta} }{ \Sigma} \left( (r^2+n^2+a^2) d{\phi} - a dt \right)^2
	+\frac{\Sigma}{\Delta}dr^2 + \Sigma d{\theta}^2,
	\label{eq_ds^2}
\end{equation}
where $\Sigma$, $\Delta$, and $\chi$ are, respectively, defined by
\begin{align}
		\Sigma &= r^2+(n+a \cos \theta)^2,\label{eq_Sig}\\
		\Delta &= r^2-2m r +a^2+q^2-n^2,\label{eq_Del}\\
		\chi &= a \sin^2 \theta - 2 n \cos \theta.\label{eq_chi}
\end{align}
Parameters $m$, $a$, $q$, and $n$ are the black hole mass, spin, electric charge, and NUT charge, respectively. The electromagnetic 2-form field is given by
\begin{align}
	F_{a b} &= \frac{q}{\Sigma^2}(r^2-(n+a \cos \theta)^2 ) (dr)_a \wedge \left( (dt)_b - \chi (d\phi)_b \right)	 \nonumber \\
	&+ \frac{2 a q r \sin \theta \cos \theta}{\Sigma^2} (d\theta)_a \wedge ( (r^2+a^2+n^2) (d\phi)_b - a (dt)_b ).
	\label{eq_F_ab}
\end{align}
Importantly, metric (\ref{eq_ds^2}) and electromagnetic field (\ref{eq_F_ab}) are solutions to the Einstein field equation, but the spacetime is not globally asymptotically flat~\cite{Misner:1963fr,manko2005physical}. From Eqs.~(\ref{eq_Sig}) and (\ref{eq_chi}), the NUT charge breaks $\mathbb{Z}_2$ symmetry defined by the transformation $\theta \rightarrow \pi - \theta$. This symmetry breaking induces that some physical quantities may be no longer mirror symmetric about the equatorial plane. Besides, when $r$ approaches infinity, the component $g_{t \phi}$ of metric will be asymptotic to $-2 n \cos \theta$, which is also a little difference from the case of the Kerr-Newman black hole. But, on the other hand, similar to Kerr-Newman black hole, Kerr-Newman Taub-NUT black hole can also have two horizons at
\begin{equation}
	r_{\pm} = m \pm \sqrt{m^2 + n^2 - a^2 - q^2},
\end{equation}
for $m^2+n^2 > a^2+q^2$. Besides, $m^2+n^2 = a^2+q^2$ corresponds to the extreme black holes. In the following sections, we will discuss both the non-extremal and extremal black holes. We write $r_+$ as $r_h$ to avoid symbol confusion with other contents.

\section{Topology for non-extremal black holes} \label{TCforBH}

Topological number are discrete, which can be used to distinguish different system structures. For a vector, its zero points, often relating with physical sources, can be treated as defects, and thus we can endow each of them with a topological charge, with which these zero points can be classified. In Refs.~\cite{Cunha:2017qtt,cunha2020stationary}, it was shown that the light rings can be cast to the zero points of the constructing vector. Following this idea, the winding number acting as topological charge of the light rings can uncover some underlying properties of the black hole solutions.

In this section, we will briefly review the equations of the light rings and study the corresponding topological charges for the black hole without $\mathbb{Z}_2$ symmetry.

\subsection{Light rings}

For a photon, its motion can be determined by the Hamiltonian
\begin{equation}
	\mathcal{H} = \frac {1}{2} g^{\mu \nu}  p_\mu p_\nu = 0,
	\label{eq_H}
\end{equation}
where $p_\mu$ is 4-momentum of the photon. Considering the black hole solution (\ref{eq_ds^2}), there are two Killing vectors, $\partial_t$ and $\partial_{\phi}$. The corresponding conserved quantities are $p_t = -E$ and $p_{\phi} = \Phi$. Combining with these results, the equation (\ref{eq_H}) can be rewritten as
\begin{eqnarray}
	T+V=0, 	\label{eq_TV}
\end{eqnarray}
where
\begin{eqnarray}
	T &=& g^{r r} p_r ^2+ g^{\theta \theta} p_{\theta}^2,
	\label{eq_T} \\
	V &=&- \frac{1}{D} (g_{\phi \phi} E^2 +2 g_{t \phi} E {\Phi}+g_{t t} {\Phi}^2),
	\label{eq_V}
\end{eqnarray}
with $D= g_{t \phi}^2 - g_{t t} g_{\phi \phi}$. For a light ring, $p_r$, $p_\theta$ and $\dot{p}_\mu$ should vanish. Therefore, photon will move around the black hole and keep $ p_a = (-E,0,0,\Phi) $ uncharged. By using Eqs.~(\ref{eq_TV}), (\ref{eq_T}), (\ref{eq_V}) and Hamiltonian canonical equation, the light rings satisfy the following conditions
\begin{equation}
	V = \partial_{r} V = \partial_{\theta} V = 0. \label{eq_pV}
\end{equation}
Moreover, the light ring is stable or unstable in $x^\mu$ direction ($\mu=r$, $\theta$) if $\partial_{\mu}^2V > 0 $ or $\partial_{\mu}^2V < 0 $. Here, we only focus on the light rings with $\Phi \neq 0$. Therefore, it is reasonable to convert formula (\ref{eq_V}) into the following form
\begin{equation}
	V = -\frac{ {\Phi}^2 g_{\phi \phi} }{ D } (E/\Phi - H_+)(E/\Phi - H_-) ,
	\label{eq_Vexpression}
\end{equation}
where $H_+$ and $H_-$ correspond to the prograde and retrograde orbits, and are given by
\begin{equation}
	H_{\pm} =\frac{-g_{t \phi}  \pm  \sqrt{D}}{g_{\phi \phi}}.
	\label{eq_Hpm}
\end{equation}
The direction of angular momentum for the prograde and retrograde photons are the same and opposite with the black hole spin. So if $a\geq0$, we shall have $\Phi>0$ for prograde case and $\Phi<0$ for the retrograde case. Without loss of generality, we only concern $a\geq0$. Employing with formula (\ref{eq_Vexpression}), the original three conditions of light rings in Eq.~(\ref{eq_pV}) can be transformed into two following conditions (for each rotating sense)
\begin{equation}
	 \partial_{r} H_\pm = \partial_{\theta} H_\pm = 0.
	 \label{eq_cond}
\end{equation}
Note that for $H_+$, the stable and unstable light rings correspond to $\partial_{\mu}^2H_+ > 0 $ and $\partial_{\mu}^2H_+ < 0 $, respectively. While it reverses for $H_-$. It seems that Eq.~(\ref{eq_cond}) can not be obtained from Eqs.~(\ref{eq_pV}) and (\ref{eq_Vexpression}) if $g_{\phi \phi} = 0$. However, this is not the case, and we show it in Appendix \ref{the light rings with gphiphi=0}. Besides, in the above discussion, we assume $D > 0$, so the conclusion here may not apply to the case that light ring coincides with or hides behind the event horizon.

\subsection{Topological charge for light rings} \label{Topolog}

For each rotating sense given in Eq.~(\ref{eq_cond}), these light rings can be well described by constructing the corresponding topological charge, or winding number. For the first step, we need to introduce a normalized field $v$
\begin{equation}
	v_r^\pm (r, \theta) = \frac{1}{\sqrt{g_{r r}(r, \theta)}} \partial_r H_\pm (r, \theta),\quad
	v_{\theta}^\pm (r, \theta)= \frac{1}{ \sqrt{g_{\theta \theta}(r, \theta) } } \partial_\theta H_\pm (r, \theta),
	\label{eq_vrt}
\end{equation}
which can be treated as a map from the physical space $X$
\begin{equation}
	X= \left\{ \left( r,\theta \right) \middle|r_{h}<r<\infty ,0<\theta <\pi \right\},
	\label{reg}
\end{equation}
to vector space $V$
\begin{equation}
	V= \left\{ \left( v_r, v_\theta \right) \middle|-\infty<v_r<\infty, -\infty<v_\theta<\infty\right\}.
\end{equation}
Space $X$ represents the physical space outside the event horizon, ensuring $D>0$ and validity of our discussion. The corresponding parameters $(r, \theta)$ of light rings in space $X$ will be mapped to the points at $(v_r, v_ \theta) = 0 $. Thus, we can use the winding number
\begin{equation}
	w = \frac{1}{2 \pi}\oint_{\tilde{C}} d \Omega
	\label{eq_w}
\end{equation}
to characterize each zero point. The angle $\Omega$ measures the direction of the vector and satisfies
\begin{gather}
	 \cos \Omega(r,\theta) = {v_r(r,\theta)}/v(r,\theta),\ \sin \Omega(r,\theta) = {v_{\theta}(r,\theta)}/v(r,\theta),
	 \label{eq_omega}
	 \\
	 v(r,\theta)^2 = v_{r}(r,\theta)^2 + v_{\theta}(r,\theta)^2.
\end{gather}
The contour $\tilde{C}$  is in space $V$, and it is given by the image of mapping from contour $C$ in space $X$. The winding number of different points can be calculated by constructing contour $C$ enclosing them:
\begin{itemize}
 \item If curve $C$ encloses a normal space point (not a light ring), the winding number is trivial.
 \item If curve $C$ encloses a zero points of vector (a light ring), the winding number is $\pm 1$. The ``+1" and ``-1" represent local extremal point and saddle point, respectively.
\end{itemize}

When we consider a large contour that encloses $N$ zero points of the vector, the integral (\ref{eq_w}) shall give the total topological number
\begin{eqnarray}
 W=\sum_{i=1}^{N}w_i.
\end{eqnarray}
If this large contour covers the whole space of $X$, $W$ will contain the contributions of all zero points of the vector $v$. As a result, it shall give the global topological property of the concerned space. Therefore, we can study the topological properties of the light rings in Taub-NUT black hole by both the winding number and topological number from local and global perspectives.

\begin{figure}
	\begin{center}
		\subfigure[  \label{Circle10}]{\includegraphics[width=6cm]{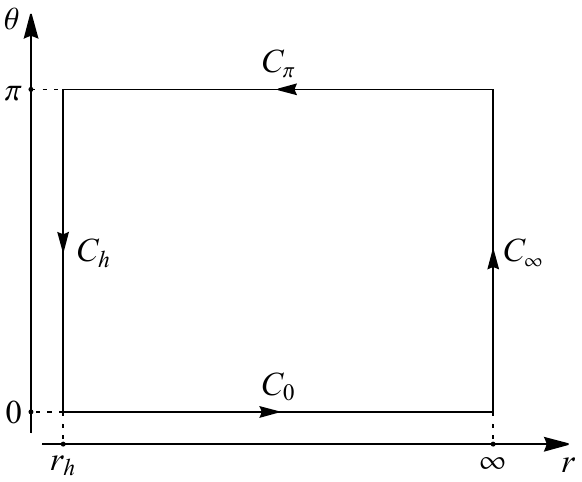}}
		\  \ \  \  \  \  \  \ \
		\subfigure[
\label{Circle11}]{\includegraphics[width=6cm]{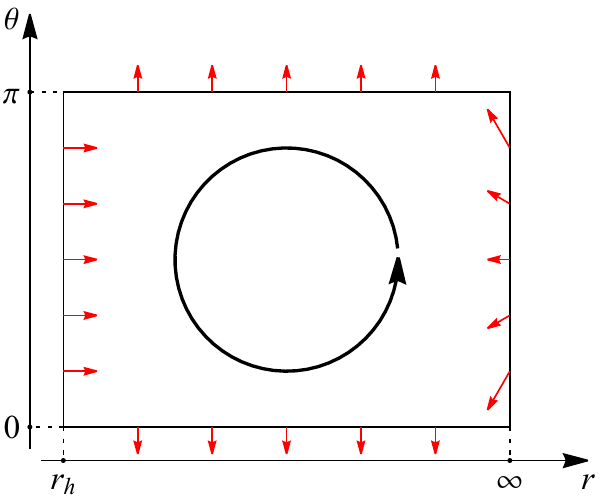}}
	\end{center}
	\caption{ Representation of the contour $C$ which encloses $X=\left\{ \left( r,\theta \right) \middle|r_{h}<r<\infty ,0<\theta <\pi \right\}$. The complete $C$ contains four segments, i.e., $C=C_0\cup C_\infty\cup C_\pi\cup C_h$. (a) The black arrows indicate the direction of the contour $C$. (b) The sketch of the direction of the vector $v$ along the contour $C$ for prograde case. The black circular arrow indicates the direction of the contour we concerned. The red arrows represent the direction of the vector.}
\end{figure}

Choosing the contour $C$ as the boundary of region $X$, the integral (\ref{eq_w}) can be decomposed as
\begin{equation}
	W =\frac{1}{2\pi}\left(\int_{C_0} d\Omega+\int_{C_\infty} d\Omega +\int_{C_\pi} d \Omega+ \int_{C_h} d\Omega \right),
	\label{eq_expandw}
\end{equation}
if the boundary is smooth. For clarity, we sketch the contour in Fig. \ref{Circle10} with $C=C_0\cup C_\infty\cup C_\pi\cup C_h$.

Next, we aim to obtain $\Delta \Omega$ by considering the asymptotic behavior of $v$ at these four segments ($C_{0, \infty, \pi, h}$). After some algebra calculations, we have following results:
\begin{itemize}
	\item Near the event horizon boundary,
	\begin{equation}
		\left. v_r^\pm \right|_{r \rightarrow  r_h}=
		\pm \frac{\sqrt{r_{h}^{2}+\left( n+a\cos \theta \right) ^2}}{\left( r_{h}^{2}+a^2+n^2 \right)^2 \sin \theta}\left(r-r_h\right)+\mathcal{O}\left(\sqrt{r-r_h}\right), 	 \quad	
		\left. v_\theta^\pm \right|_{r \rightarrow  r_h} = \mathcal{O}\left(\sqrt{r-r_h}\right).
		\label{eq_vh}
	\end{equation}
	\item Near the infinity boundary,
	\begin{equation}
		\left. v_r^\pm \right|_{r \rightarrow   \infty}=\mp \frac{1}{r^2 \sin \theta}+\mathcal{O}\left(1/r^3\right),\quad
		\left. v_\theta^\pm \right|_{r \rightarrow   \infty} = \mp \frac{\cos \theta}{r^2 \sin^2 \theta }+\mathcal{O}\left(1/r^3\right).
		\label{eq_vi}
	\end{equation}
	\item Near the $\theta = 0$ boundary,
	\begin{equation}
		\left. v_r^\pm \right|_{\theta \rightarrow  0} = \mathcal{O}\left(\theta \right),\quad
		\left. v_\theta^\pm \right|_{\theta \rightarrow  0}=\mp \frac{\sqrt{r^2+(a+n)^2}}{4 n^2 \sqrt{r^2 - 2 mr+a^2+q^2-n^2}}+\mathcal{O}\left(\theta \right).
		\label{eq_v0}
	\end{equation}
	\item Near the $\theta = \pi$ boundary,
	\begin{equation}
		\left. v_r^\pm \right|_{\theta \rightarrow  \pi} = \mathcal{O}\left(\pi-\theta\right),\quad
		\left. v_\theta^\pm \right|_{\theta \rightarrow  \pi}=\pm \frac{\sqrt{r^2+(a-n)^2}}{4 n^2 \sqrt{r^2 - 2 mr+a^2+q^2-n^2}}+\mathcal{O}\left(\pi-\theta\right).
		\label{eq_vp}
	\end{equation}
\end{itemize}
From above analysis, we can find the direction of the vector on these four segments. For the prograde orbit case marked with ``+", $v$ is downward, left, up, and right for $C_{0, \infty, \pi, h}$. While for the retrograde case marked with ``-", the direction of the vector reveres. Nevertheless, their topological number keeps the same
\begin{equation}
	    W^\pm = -1,
\end{equation}
which means that there is at least one unstable light ring outside the black hole horizon. As a supplement, the sketch of boundary vector field for prograde light rings is shown in Fig. \ref{Circle11}. It is worth noting that when $n \rightarrow \infty$, the values of (\ref{eq_v0}) and (\ref{eq_vp}) seem to become imaginary. However, this is not true because $r$ must be greater than $r_h$, which ensures that the formula in the square root is positive. Further comparing with the Kerr-Newman black hole, a direct conclusion is that although the NUT charge $n$ affects the vector field $v$ on the boundary, the topological number corresponding to the light rings does not change. We also must be careful that the current discussion is only valid for non-extreme black hole, which is because that the leading term of $\left. v_r^\pm \right|_{r \rightarrow  r_h}$ will change for the extremal black hole. And more specific analysis will be shown in section \ref{Extreme BH}.

\begin{figure}
	\begin{center}
		\subfigure[ \label{Circle20}]{\includegraphics[width=7cm]{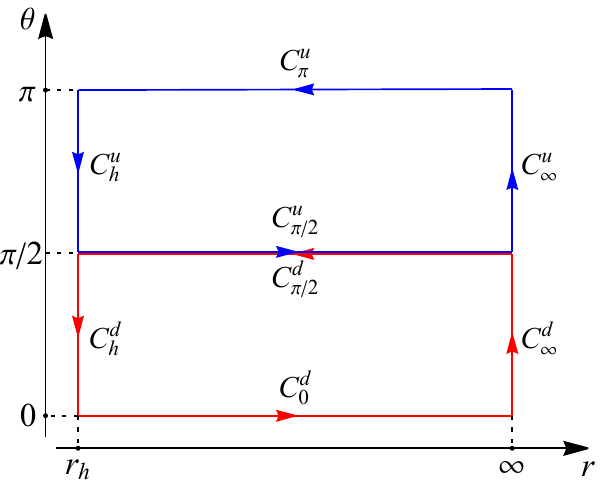}}
		\  \
		\subfigure[
		 \label{Circle21}]{\includegraphics[width=7cm]{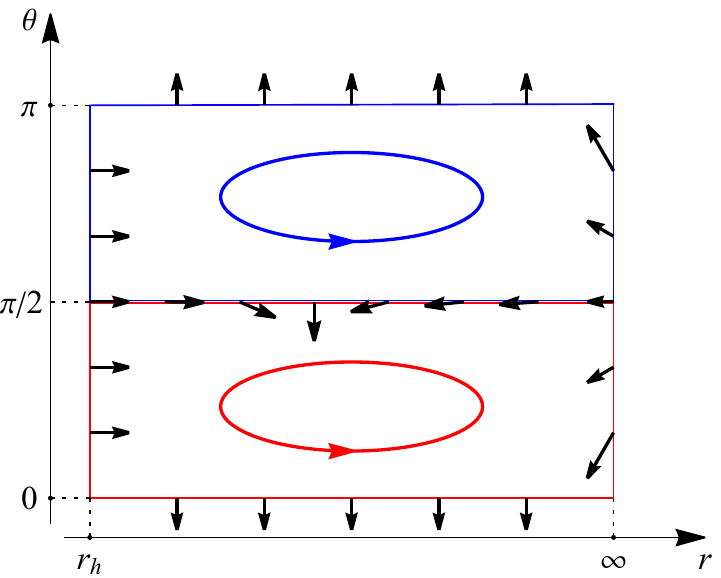}}
\end{center}
\caption{Representations of the contour $C^u$ and $C^d$ which encloses $X_u=\left\{ \left( r,\theta \right) \middle|r_{h}<r<\infty ,\pi/2\leqslant\theta <\pi \right\}$ and $X_d=\left\{ \left( r,\theta \right) \middle|r_{h}<r<\infty, 0<\theta \leqslant\pi/2 \right\}$, respectively. Four blue line segments (upper side of $\theta=\pi/2$) represent curve $C^u=C^{u}_{\pi/2}\cup C^{u}_{\infty}\cup C^{u}_{\pi}\cup C^{u}_{h}$ and four red line segments (under side of $\theta=\pi/2$) represent curve $C^d=C^{d}_{\pi/2}\cup C^{d}_{h}\cup C^{d}_{0}\cup C^{d}_{\infty}$. (a) The arrows indicate the direction of contours $C^u$ and $C^d$. (b) The sketch of vector direction  along these line segments for prograde light rings in the case of $n>0$ and $a>0$. The red and blue circular arrows indicate the directions of the contours $C^u$ and $C^d$, respectively. The black arrows indicate the direction of vector along the boundary.}
\end{figure}

Reviewing the line element (\ref{eq_ds^2}), $\mathbb{Z}_2$ symmetry of $\theta \rightarrow \pi - \theta$ is breaking for the Kerr-Newman Taub-NUT black hole. But it fails to affect the topological number of the vector $v$, leading to the fact that Kerr-Newman Taub-NUT black hole and Kerr-Newman black hole are in the same topological class for the light rings. This is mainly because that the topological number represents a global property rather than a local one. In order to uncover the influence of $\mathbb{Z}_2$ symmetry on the topological number, we define following two contour integrals
\begin{align}
	W_u &=\frac{1}{2\pi} \int_{C^u} d \Omega=\frac{1}{2\pi}\left( \int_{C_{\pi/2}^u} d \Omega+\int_{C_\infty^u} d \Omega
	+\int_{C_\pi^u} d \Omega
	+\int_{C_h^u} d \Omega\right),
	\label{eq_wu}
	\\
	W_d &=\frac{1}{2\pi} \int_{C^d} d \Omega=\frac{1}{2\pi}\left( \int_{C_0^d} d \Omega+\int_{C_\infty^d} d \Omega+\int_{C_{\pi/2}^d} d \Omega+\int_{C_h^d} d \Omega \right).
	\label{eq_wd}
\end{align}
The contours $C^u$ and $C^d$ are shown in Fig. \ref{Circle20}, which are for the upper half-plane and lower half-plane, respectively. The corresponding topological numbers $W_u$ and $W_d$ are for the regions
\begin{align}
	X_u&=\{ (r,\theta ) |  r_h<r<\infty, \pi/2\leqslant\theta<\pi \},
	\\
	X_d&=\{ (r,\theta ) |  r_h<r<\infty, 0<\theta\leqslant\pi/2 \},
\end{align}
respectively. This further requires us to consider the behavior of vector $v$ on the equatorial plane with $\theta = \pi/2$, which are
\begin{align}
	\left. v_r^\pm \right|_{\theta \rightarrow \pi /2}&=\pm \frac{(r-m)\left( r^2+n^2 \right) -2 r \sqrt{\Delta} \left( \pm a + \sqrt{\Delta} \right)}{\sqrt{r^2+n^2}\left( r^2+n^2+a^2 \pm a\sqrt{\Delta} \right) ^2}+\mathcal{O} \left( \theta -\frac{\pi}{2} \right),
	\label{eq_vr2}
	\\
	\left. v_{\theta}^\pm \right|_{\theta \rightarrow \pi /2}&=- \frac{2n\sqrt{\Delta}\left( \pm a+\sqrt{\Delta} \right)}{\sqrt{r^2+n^2}\left( r^2+n^2+a^2 \pm a\sqrt{\Delta} \right) ^2}+\mathcal{O} \left( \theta -\frac{\pi}{2} \right)
	\label{eq_vt2},
\end{align}
for $n \ne 0$. When $n$ vanishes, Eq.~(\ref{eq_vt2}) reduces to
\begin{equation}
	\left. v_{\theta}^\pm \right|_{\theta \rightarrow \pi /2}=\pm \frac{\sqrt{\Delta}\left( r^2+2a^2 \pm 2a\sqrt{\Delta} \right)}{2r\left( r^2+a^2 \pm a\sqrt{\Delta} \right) ^2}(\theta -\frac{\pi}{2})+\mathcal{O} \left( (\theta -\frac{\pi}{2})^2 \right).
\end{equation}
By using $r>m$ and $m^2 > a^2+q^2$, it is easy to find $ r^2+2a^2\pm2a\sqrt{\Delta} > 0$.

To clarify this issue, we take the prograde light ring and NUT charge $n>0$ as an example. Other cases would be the similar. Note that in both prograde and retrograde cases, the black hole spin $a$ is set to be positive in priority. In Fig. \ref{Circle21}, we show the direction of the vector $v^+$ on the segments of $X_{u}$ and $X_{d}$, where the negative $v_\theta^+$ has been considered. Accordingly, we easily obtain $W_{u}=-1$ and $W_{d}=0$. Besides, since $C=C^{u}\cup C^{d}$, one easily gets $W=W_{u}+W_{d}$.

In other cases, we summarize the results in Table \ref{T1}. Note that when $n$ vanishes, it reduces to the Kerr-Newman case, and our above analysis fails because its light ring always on the equatorial plane, while we list the total topological number for a comparison. From the table, it is clear that $W_d$ and $W_u$ take different values for each case. However their sum is always -1, further confirming the relation $W=W_d+W_u$. This result also indicates that the unstable light ring is shifted off the equatorial plane by the NUT charge, but the number of the light ring keeps unchanged. Therefore, the NUT charge does not alter the topological class of the light ring even when the $\mathbb{Z}_2$ symmetry is broken.

To provide an more intuitive understanding, we give some examples on the behaviors of the vector $v$ in $X$ space in Appendix \ref{Figure of V F}. The deflection angle $\Delta\Omega$ is also shown, from which one can easily obtains the value of the topological number corresponding to the light rings.

\begin{table}
	\centering
	\resizebox{\linewidth}{!}{
	\begin{tabular}{|c|c|c|c|}
		\hline
		``$+$"& \  $X_d=\{ (r,\theta ) | r_h<r<\infty, 0<\theta\leqslant\pi/2 \}$ \ & \ $X_u=\{ (r,\theta ) | r_h<r<\infty, \pi/2\leqslant\theta<\pi \}$  \ & \ $X=\{ (r,\theta ) | r_h<r<\infty, 0<\theta<\pi \}$  \\
		\hline		
		\ $n>0$ \ & $W_d =0$ & $W_u=-1$ & $W=-1$ \\
		\hline		
		\ $n<0$ \ & $W_d=-1$ & $W_u=0$ & $W=-1$ \\
		\hline
		\ $n=0$ \ & - & - & $W=-1$ \\
		\hline
		``$-$"& \  $X_d=\{ (r,\theta ) | r_h<r<\infty, 0<\theta\leqslant\pi/2 \}$ \ & \ $X_u=\{ (r,\theta ) | r_h<r<\infty, \pi/2\leqslant\theta<\pi \}$  \ & \ $X=\{ (r,\theta ) | r_h<r<\infty, 0<\theta<\pi \}$  \\
		\hline		
		\ $n>0$ \ & $W_d =-1$ & $W_u=0$ & $W=-1$ \\
		\hline		
		\ $n<0$ \ & $W_d=0$ & $W_u=-1$ & $W=-1$ \\
		\hline
		\ $n=0$ \ & - & - & $W=-1$ \\
		\hline
	\end{tabular}}
	\caption{Topological numbers corresponding to different regions for prograde (``+") and retrograde (``-") light rings.}
	\label{T1}
\end{table}

\section{Solutions of light rings}\label{Solution}

In the previous section, one can draw a conclusion that Kerr-Newman Taub-NUT black hole has at least one light ring. Due to the NUT charge, the light ring will deviate from the equatorial plane. On the other hand, there may be a possibility that the contribution of a saddle point and a extremal point to the topological number cancels, so the specific number of the light ring could not be uniquely determined by the topological number. In order to show the detailed information of the light ring, we shall examine the zero point of the vector $v$.

The equation of $v=0$ for prograde light ring can be reduced to
\begin{align}
	\partial _r H_+&=\frac{f\left( r,\theta \right) \partial _r\Delta \sin \theta}{2\sqrt{\Delta}\left( \left( r^2+a^2+n^2 \right) \sin \theta +\left( a\sin ^2\theta -2n\cos \theta \right) \sqrt{\Delta} \right) ^2}=0,
	\label{eq_vp1} \\
	\partial _{\theta}H_+&=\frac{-g\left( r,\theta \right) \sqrt{\Delta}}{\left( \left( r^2+a^2+n^2 \right) \sin \theta +\left( a\sin ^2\theta -2n\cos \theta \right) \sqrt{\Delta} \right) ^2}=0,
	\label{eq_vp2}
\end{align}
where
\begin{gather}
	f \left(r,\theta \right) = -	4  r  \frac{ (a \sin \theta  + \sqrt{\Delta }) \sqrt{\Delta } }{\partial_r \Delta } +	r^2 + ( n+a \cos \theta )^2,
	\label{eq_f}\\
	g\left(r,\theta \right) = -a^2 \cos ^3\theta  +  \left(r^2 + n^2+ 2 a^2 \right) \cos \theta +2 a n + 2 \left(n + a \cos \theta \right) \sin \theta  \sqrt{\Delta} \ .
	\label{eq_g}
\end{gather}
The retrograde case is not shown, which can be obtained by some relations with the prograde case. For example, changing both sign of $a$ and $n$ in Eqs.~(\ref{eq_vp1}) and (\ref{eq_vp2}), one shall obtain these corresponding equations for the retrograde case. So for simplicity, we only focus on the prograde case marked with ``$+$". In addition, the black hole discussed in this section is non-extremal, and the region we concerned is $X=\left\{ (r,\theta)|r_h<r<\infty, 0<\theta<\pi \right\} $. Note that, Eqs. (\ref{eq_vp1}) and (\ref{eq_vp2}) is equivalent to $f(r,\theta) = 0$ and $g(r,\theta) = 0$. Now let us turn to the discussion about the number of roots for Eqs. (\ref{eq_vp1}) and (\ref{eq_vp2}). First, Eq.~(\ref{eq_vp1}) gives the equation $f \left(r,\theta \right) =0$. Here we attempt to examine the partial derivative of $f \left(r,\theta \right) $
\begin{equation}
	\partial_{r}  f \left(r,\theta \right) = - \frac{2\left( r(r-m)^2-m\Delta \right) \left( a\sin \theta +\sqrt{\Delta} \right)}{(r-m)^2\sqrt{\Delta}}.
\end{equation}
For the case of $a \geqslant 0$, we always have $\partial_{r}  f \left(r,\theta \right) < 0$, thus $ f \left(r,\theta \right)$ is monotonically decreasing with $r$. While when $a < 0 $, there will be a root $r=r_0$ outside the event horizon such that $\left.\partial_r f \left(r,\theta \right) \right|_{r_h<r <r_0}  > 0$ and $\left.\partial_r f \left(r,\theta \right) \right|_{r_0<r <\infty}  < 0$. Thus, $ f \left(r,\theta \right) $ increases with $r$ in interval $(r_h,r_0) $ and decreases in interval $(r_0,\infty) $. Combining with these boundary behaviors
\begin{equation}
	f \left(r_h,\theta \right) =  r_h^2 + ( n+a \cos \theta )^2 > 0, \	 f \left(\infty,\theta \right) = \lim_{r \rightarrow \infty} -r^2 < 0,
\end{equation}
and the continuity of function, we arrive an interesting conclusion: for any $0<\theta< \pi$, there is only one root $\hat{r} \in (r_h,\infty)$ such that $f(\hat{r},\theta) = 0$ and $\left.\partial_{r} f(r,\theta)\right|_{r=\hat{r}} < 0$ hold. For the convenience of discussion, we denote $r = r(\theta) $ as the solution of $f(r,\theta)=0$. Then, we turn to the remaining Eq.~(\ref{eq_vp2}). Substituting $r=r(\theta)$ into it, we obtain
\begin{equation}
	\left. \partial_\theta H_+\left ( r, \theta \right) \right|_{r=r(\theta)} = 0,
	\label{eq_Hrtheta}
\end{equation}
which is an equation with only one variable $\theta$. Solving it, we shall obtain the light rings determined by Eqs.~(\ref{eq_vp1}) and (\ref{eq_vp2}). To explore the root of Eq.~(\ref{eq_Hrtheta}), we focus on
\begin{align}
	\frac{d}{d\theta}\left(	\left. \partial_\theta H_+\left ( r, \theta \right) \right|_{r=r(\theta)} \right) &=\left.\frac{dr}{d\theta}\partial _r\partial _{\theta}H_+ \right|_{r=r(\theta)}+\left.\partial _{\theta}^{2}H_+\right|_{r = r(\theta)}
	\\
	&=\left. \frac{1}{\partial _{r}^{2}H_+\left( r,\theta \right)}\left( -\left( \partial _r\partial _{\theta}H_+ \right) ^2+\partial _{r}^{2}H_+\left( r,\theta \right) \partial _{\theta}^{2}H_+\left( r,\theta \right) \right) \right|_{f(r,\theta) = 0}.
	\label{eq_Hf}
\end{align}
If one substitutes the form (\ref{eq_Hpm}) of $H_+$, this equation will become very complex and hard to analyze. However, we have
\begin{equation}
	\left. \frac{1}{\partial _{r}^{2}H_+\left( r,\theta \right)}\left( -\left( \partial _r\partial _{\theta}H_+ \right) ^2+\partial _{r}^{2}H_+\left( r,\theta \right) \partial _{\theta}^{2}H_+\left( r,\theta \right) \right) \right|_{f(r,\theta) = 0,\ g(r,\theta)=0} >0,
	\label{eq_Hfg}
\end{equation}
for the light rings. This is mainly because that, with the conditions (\ref{eq_vp1}) and (\ref{eq_vp2}) or $f(r,\theta) = 0$ and $g(r,\theta) = 0$, we have
\begin{equation}
	\partial_r^2 H_+ \propto \partial_r f(r,\theta) < 0,  \ \partial_\theta^2 H_+ \propto - \partial_\theta g(r,\theta) >0,
	\label{eq_ufg}
\end{equation}
where $ \partial_\theta g(r,\theta) <0$ can be found in Appendix \ref{proof2}. It must be noted that compared to formula (\ref{eq_Hf}), Eq.~(\ref{eq_Hfg}) has an additional condition $g(r,\theta)=0$. That means that Eq.~(\ref{eq_Hfg}) is the behavior of formula (\ref{eq_Hf}) at any zero points of $g(r,\theta)=0$. Thus, based on the zero point theorem of continuous functions and
\begin{equation}
		\left. \partial_\theta H_+\left ( r, \theta \right) \right|_{r=r(\theta = 0),\theta = 0} <0,  \
		\left. \partial_\theta H_+\left ( r, \theta \right) \right|_{r=r(\theta = \pi),\theta = \pi} >0,
\end{equation}
Eq.~(\ref{eq_Hf}) is positive at any zero points of the Eq.~(\ref{eq_Hrtheta}), implying that it has only one root. Therefore, we arrive a conclusion: Kerr-Newman Taub-NUT black hole has only one prograde light ring. Moreover, employing with the relation between the prograde and retrograde cases, this conclusion is also valid for the retrograde light ring. Besides, due to inequality (\ref{eq_ufg}), these light rings are stable in $\theta$ direction and unstable in $r$ direction. This means that the light ring corresponds to the saddle point, which is consistent with the result of winding number.

\begin{figure}
	\begin{center}
		\subfigure[ \label{n_theta}]{\includegraphics[width=7cm]{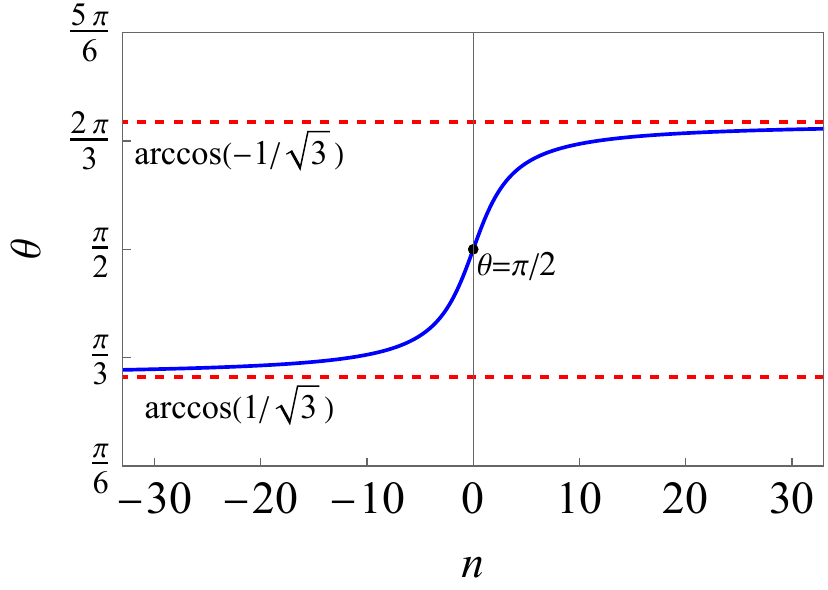}}
		\  \
		\subfigure[
		 \label{n_rthetapdf}]{\includegraphics[width=7.1cm]{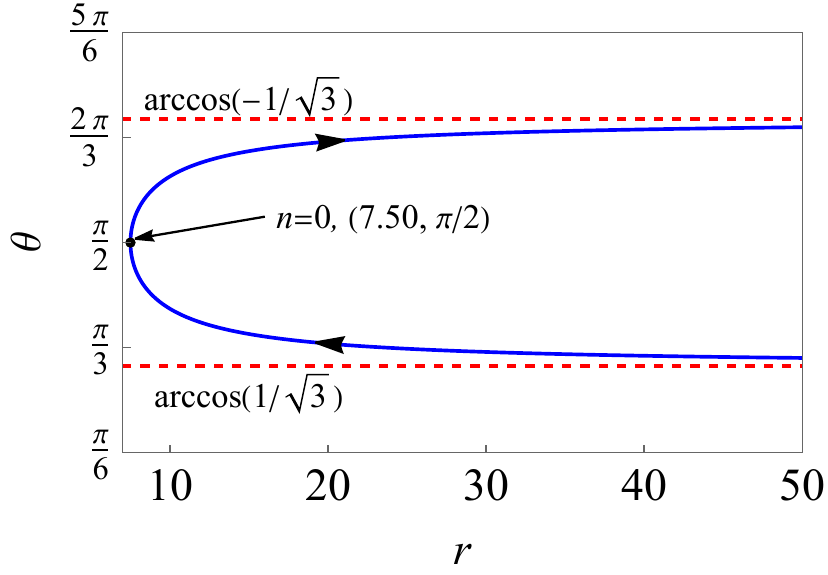}}
		\  \
	\end{center}
	\caption{Numerical solutions for prograde light ring with $m=3$, $q=1$, $a=1$. The blue curves are for the light rings, and the red dashed lines represent $\theta = \arccos(-1/\sqrt 3)$ and $\theta = \arccos(1/\sqrt 3)$. (a) The angular location of the light rings as a function of the NUT change $n$. (b) The location of the light rings in $r$-$\theta$ plane with varying $n$. The arrows represent the increasing $n$ from $-\infty$ to $\infty$. In addition, (7.50, $\pi/2$) denotes the light ring when $n$ vanishes.}
	\label{n_rtheta}
\end{figure}

Although we know from the above analysis that there is only one light ring outside the black hole horizon, the analytical solution is still hard to obtain. However, when we consider the case of $n \rightarrow \pm\infty$, the analytical solution reads
\begin{equation}
	r = \sqrt{3} n, \ \theta =  \arccos \left(\mp \frac{1}{\sqrt{3}}\right).
\end{equation}
Moreover, we also numerically solve the light rings, and exhibit them in Fig. \ref{n_rtheta} for the prograde light rings. From Fig. \ref{n_theta}, it is easy to see that there always exists one and only one light ring for an arbitrary value of $n$. The deviation angle is also bounded by $\mp \arccos(1/\sqrt{3})$. With the increase of NUT charge $n$, we also display the light ring in Fig. \ref{n_rthetapdf}. The asymptotic behaviors of $\theta$ at large $n$ are also explicitly shown. Meanwhile, in the separated ranges ($0$, $\pi/2$) and ($\pi/2$, $\pi$), $\theta$ is a monotonic function with the radial radius of the light ring. This result is also consistent with our above analysis. From the perspective of topology, we always have a conserved topological number $W=-1$ indicating the existence of an unstable light ring. This result is also independent of the black hole parameters. So, one can obtain some interesting properties of the light ring without solving the corresponding equations.

\section{Topology for extremal black holes}
\label{Extreme BH}

In above sections, we have studied the topological properties of the light rings for a non-extremal Kerr-Newman Taub-NUT black hole. However, for an extremal black hole, the topological properties is hard to study due to the fact that the behavior of vector $v$ near the degenerate event horizon is relatively dependent on the specific metric. Therefore, exploring the corresponding topology has potential enlightening significance on understanding the light rings for the extremal black holes. Recent study on this issue in spherically symmetric black hole backgrounds can be found in Refs.~\cite{Hod:2022EBH_LR,Bargueno:2022EBL_S_LR,Peng:2022EFHBH_NCG}.

\begin{figure}
	\begin{center}
		\subfigure[
		 \label{Circle31}]{\includegraphics[width=7cm]{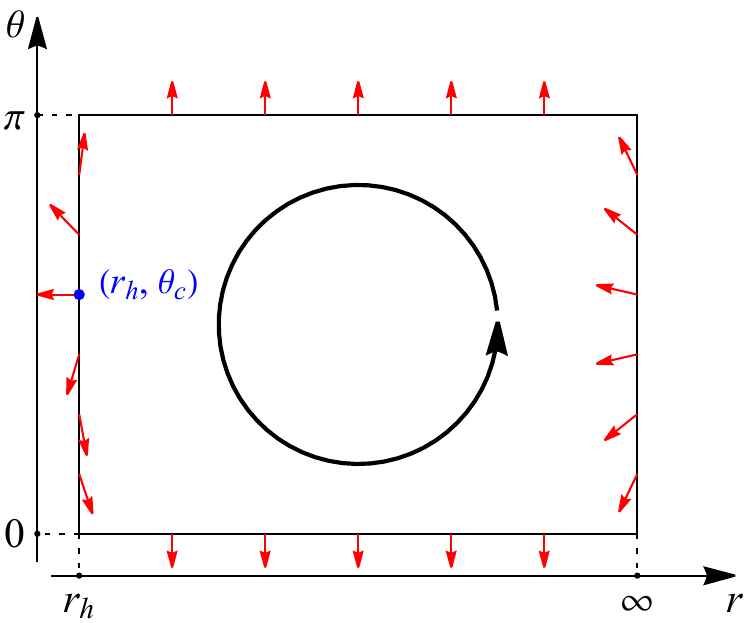}}
		\  \
		\subfigure[ \label{Circle32}]{\includegraphics[width=7cm]{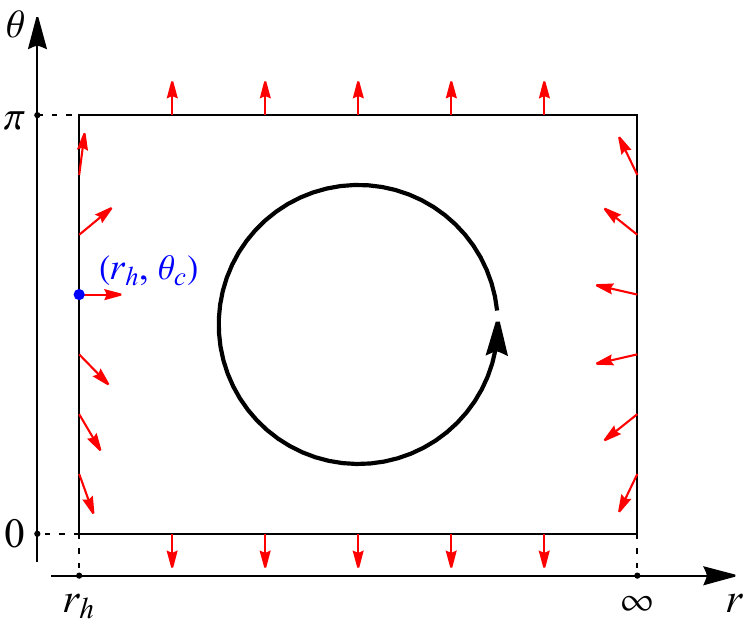}}
	\end{center}
	\caption{The sketch of vector on the boundary for prograde case. Four black line segments represent the contour $C$ which is the boundary of $X=\left\{ \left( r,\theta \right) \middle|r_{h}<r<\infty ,0<\theta <\pi \right\}$. The black circular arrows indicate the directions of contours. The red arrows represent the direction of vector on the boundary. ``$\theta_c$" in blue color of ($r_h$, $\theta_c$) corresponds is given in formula (\ref{eq_theta_0}). (a) Case for $V_c<0$.  (b) Case for $V_c>0$.}
\end{figure}

Now, we attempt to examine the topology of light rings for the extremal spinning Kerr-Newman Taub-NUT black hole. The metric function $\Delta$ reads
\begin{equation}
	\Delta = \left( r - m \right)^2,
\end{equation}
due to the constraint $m^2+n^2=a^2+q^2$. This means that these two horizons of the black hole coincide exactly at $r_h = r_+ = r_-=m$. Thus, the vector $v$ (\ref{eq_vrt}) near the horizon shall be significantly changed, whereas slightly modified on other boundaries.  For clarity, we show the vector behavior near the horizon
\begin{align}
		\left. v_{r}^{\pm} \right|_{r\rightarrow r_h}&=\pm \frac{\left( n+a\cos \theta \right) ^2+m^2\mp 2ma\sin \theta}{\sin \theta \left( m^2+n^2+a^2 \right) ^2\sqrt{m^2+\left( n+a\cos \theta \right) ^2}}(r-r_h)+\mathcal{O} \left( (r-r_h)^2 \right) ,
		\label{eq_veh_r}
		\\
		\left. v_{\theta}^{\pm} \right|_{r\rightarrow r_h}&=\pm \frac{a^2\cos ^3\theta -\left( m^2+n^2+2a^2 \right) \cos \theta -2an}{\sin ^2\theta \left( m^2+n^2+a^2 \right) ^2\sqrt{m^2+\left( n+a\cos \theta \right) ^2}}(r-r_h)+\mathcal{O} \left( (r-r_h)^2 \right)
		\label{eq_veh_theta}.
\end{align}
Since the concerned region is outside the black hole horizon, we need to examine the leading terms of Eqs.~(\ref{eq_veh_r}) and (\ref{eq_veh_theta}) near the event horizon. Although these equations are more complicated than that of the non-extremal black holes, we are fortunate enough to solve $v_\theta^\pm = 0$ to the first order as $\theta = \theta_c$ given by
\begin{equation}
	\cos \theta_c =-\frac{2\sqrt{\left( m^2+n^2+2a^2 \right)}}{\sqrt{3}a}\sin \left( \frac{\pi}{6}-\frac{1}{3}\arccos \left( \frac{3\sqrt{3}a^2n}{\sqrt{\left( m^2+n^2+2a^2 \right) ^3}} \right) \right) .
	\label{eq_theta_0}
\end{equation}
The simple result indicates that, $\left.v_\theta^+ \right|_{r \rightarrow r_h}<0$ and $\left.v_\theta^- \right|_{r \rightarrow r_h}>0$ for $0<\theta<\theta_c$. If $\theta_c<\theta<\pi$, their values reverse. Further combining with the fact that $\left. v_{\theta}^+ \right|_{\theta \rightarrow 0}<0$, $ \left. v_{\theta}^+ \right|_{\theta \rightarrow \pi}>0$, $\left. v_{\theta}^- \right|_{\theta \rightarrow 0}>0$, and $ \left. v_{\theta}^- \right|_{\theta \rightarrow \pi}<0$, we find
\begin{equation}
	V_c^\pm \equiv \text{the leading term of} \ \left. v_{r}^\pm \right|_{\theta = \theta_c, r\rightarrow r_h}
\end{equation}
is critical to the topological number,
\begin{equation}
	\Delta \Omega_h=\int_{C_h} d\Omega = \mp \pi \sgn \left( V_{c}^{\pm} \right) .
	\label{eq_ChVOmega}
\end{equation}
Counting the contributions from other segments of the boundary, we reach the following conclusion:
\begin{itemize}
	\item For the prograde case, the topological number is 0 for $V_c^+ < 0$ (shown in Fig. \ref{Circle31}), and -1 for $V_c^+>0$ (shown in Fig. \ref{Circle32}).
	\item For the retrograde case, the topological number is -1 and 0 for  $V_c^- < 0$ and  $V_c^- > 0$, respectively.
\end{itemize}

Therefore, quite different from the non-extremal black holes, all the extremal black hole can not be classified into the same topological class simply. In particular, as shown above, when $V_c^\pm$ changes its sign, a topological phase transition must occur. Some examples will be given in the following contents, and some supplementary materials can also be found in Appendix \ref{Example3}.

It is also worth to note that outside the black hole horizon, the leading term of $V_c^ \pm$ does not vanish for a spinning black hole, which can ensure that $\Delta \Omega_h$ for horizon boundary is clear and formula (\ref{eq_ChVOmega}) is correct. The reason is given in followings. Let us focus on the expansion
\begin{equation}
	\left. v_{r}^{\pm} \right|_{\theta =\theta _c,r\rightarrow r_h} = V_c^{\pm(1)} (r - r_h) + V_c^{\pm(2)}(r - r_h)^2 +\mathcal{O}\left( (r - r_h)^3\right),
\end{equation}
where $V_c^{\pm (1)}$ are
\begin{equation}
	V_c^{\pm (1)} = \pm \frac{\left( n+a\cos \theta _c \right) ^2+m^2\mp 2ma\sin \theta _c}{\sin \theta _c\left( m^2+n^2+a^2 \right) ^2\sqrt{m^2+\left( n+a\cos \theta _c \right) ^2}}.
	\label{eq_Vc1}
\end{equation}
When $V_c^{+ (1)}$ vanishes, one has $a>0$ and $V_c^+=V_c^{+ (2)}\propto - a<0$, which ensures that $V_c^+$ does not vanish. In the same way, $V_c^{-}$ can also be analyzed. Thus, the topological number can be obtained and expressed as
\begin{equation}
	W^{\pm}=\left\{ \begin{matrix}
		-\frac{1}{2}\left( 1\pm \sgn \left( V_{c}^{\pm \left( 1 \right)} \right) \right) 	 ,&		\text{if}\,\, V_{c}^{\pm \left( 1 \right)}\ne 0.\\
		0,&		\text{if}\,\, V_{c}^{\pm \left( 1 \right)}=0.\\
	\end{matrix} \right.
\end{equation}
If $a>0$ is fixed, it can be found that $V_c^{-(1)}$ is always less than zero, which means that the retrograde light rings always exist; but for $V_c^{+(1)}$, it is not always greater than zero, implying that prograde light ring might not exist.

\subsection{Example 1: Extremal Kerr-Newman black holes}\label{Example1}

If we take $n = 0$,  the solution will reduce to the extremal Kerr-Newman black hole with $m^2 = a^2+ q^2$. Eq.~(\ref{eq_theta_0}) will be greatly simplified and it gives $\theta_c = \pi/2$. Therefore, the functions $V_c^{\pm (1)}$ are
\begin{equation}
	V_c^{\pm (1)}= \pm \frac{(m \mp 2a)}{\left( m^2+a^2 \right) ^2}.
\end{equation}
Employing with this result, we can obtain the corresponding topological number for different black hole spin. The results are summarized in Table \ref{T.ExK_NBH}. For the prograde light ring, we see that there exists a topological phase transition at $a=m/2$, below or above which the topological number $W$=-1 or 0, respectively. This indicates that there exists one unstable prograde light ring for the lowly spinning extremal Kerr-Newman black holes, while no for the rapidly spinning black holes. Of particular interest is that there always exists one retrograde light ring for the extremal Kerr-Newman black holes. So we conclude that the topological phase transition exists for the prograde light ring, while absent for the retrograde light ring.

\begin{table}
	\centering
	\resizebox{12cm}{!}{
		\begin{tabular}{ccccc}
			\hline
			 \ \ \ \ \ \ Range \ \ \ \ \ \ & \ \ \ \ \ \ $V_c^+$ \ \ \ \ \ \ &  \ \ \ \ \ \ ``$+$" direction\ \ \ \ \ \ & \ \ \ \ \ \ $V_c^-$ \ \ \ \ \ \ & \ \ \ \ \ \ ``$-$" direction \ \ \ \ \ \ \\
			\hline
			$ 0<a < m/2 $  & $V_c^+>0$ & $W=-1$ & $V_c^-<0$ & $W =- 1$  \\
			$a \geqslant m/2 $ \ \ & $V_c^+<0$ & $W=0$ & $V_c^-<0$ & $W =- 1$ \\
			\hline
	\end{tabular}}
	\caption{Topological number for prograde and retrograde light rings outside the extreme Kerr-Newman black hole.}
	\label{T.ExK_NBH}
\end{table}

On the other hand, for the extreme Kerr-Newman black hole, the location of prograde light rings is determined
\begin{eqnarray}
	a^4\cos ^2\theta -r(2a\sin \theta -2m+r)=0,\label{eadada}\\
\left( r^2+a^2+a^2\left( 1-\cos \theta \right) +2a\sin \theta (r-m) \right) \cos \theta =0,\label{eadadab}
\end{eqnarray}
solving which gives
\begin{equation}
	\theta^+_{LR}= \pi/2,\quad r^+_{LR} = 2(m-a).
\end{equation}
Since we require a light ring outside the black hole horizon $r^+_{LR} >m$, this exactly gives our above result that the prograde light ring only exists for $a<m/2$.

For the retrograde case, we only need to change $a$ to $-a$ in Eqs.~(\ref{eadada}) and (\ref{eadadab}). The solution is
\begin{equation}
	\theta^-_{LR}= \pi/2, \quad r^-_{LR} = 2(m+a).
\end{equation}
Obviously, the radius of the retrograde light rings is always larger than the radius of the extremal black hole horizon. So one is expected to get topological number $W=-1$ regardless of the black hole spin.

\subsection{Example 2: Extremal Kerr-Newman Taub-NUT black holes}\label{Example2}

Now, let us turn to the case with $n \ne 0$. For the extremal Kerr-Newman Taub-NUT black hole, one requires $m^2 + n^2 = a^2+q^2$ on these black hole parameters. Without loss of generality, we take $m = 3$ and $n = 1$ as an example. Then we have $0 \leqslant a =\sqrt{10-n^2} \leqslant \sqrt{10}$.

\begin{figure}
	\begin{center} \label{a_Vc}{\includegraphics[width=9cm]{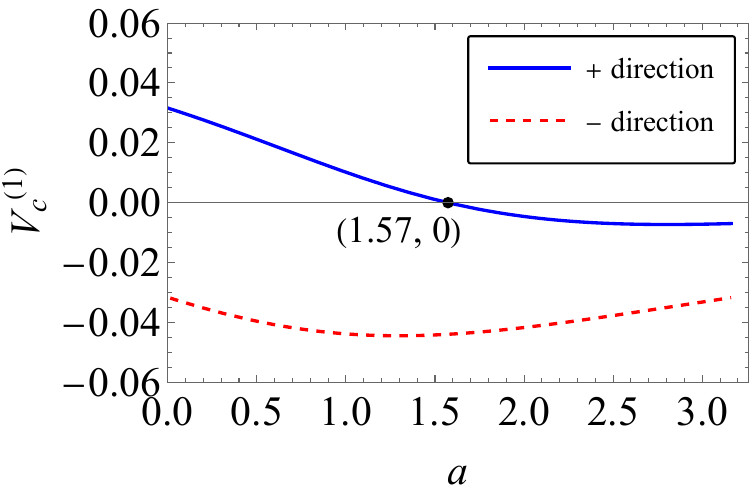}}
	\end{center}
	\caption{Quantities $V_c^{\pm(1)}$ as a function of black hole spin $a$ with $m=3$ and $n =1$.}
	\label{FIG.a_Vc1}
\end{figure}

For simplicity, we plot the first order term $V_c^{\pm (1)}$ in Fig. \ref{FIG.a_Vc1} for both prograde and retrograde cases. It is clear that $V_c^{+(1)}$ changes its sign at $a\approx1.57$. So there exists a topological phase transition for the prograde light ring. Below and above that value, there exists one or no light ring, and the topological number $W$ changes from -1 to 0. While for the retrograde case, $V_c^{-(1)}$ always takes negative values. So no topological phase transition occurs, and $W=-1$ for an arbitrary black hole spin $a$. This results are also consistent with that of Kerr-Newman black hole.

\begin{figure}
	\begin{center}
		\subfigure[
\label{rtheta1+}]{\includegraphics[width=7cm]{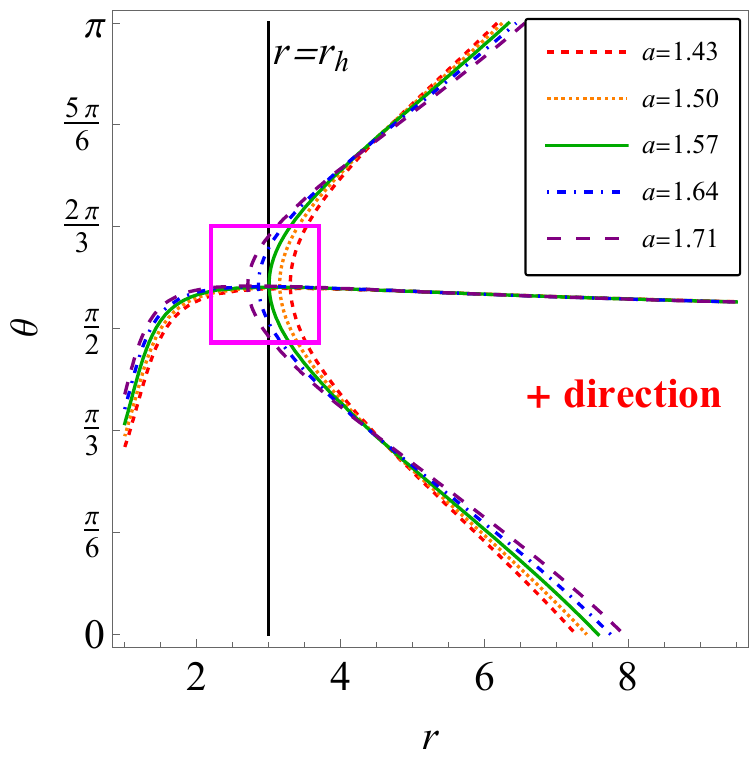}}
		\  \  \
		\subfigure[ \label{rtheta2+}]{\includegraphics[width=7cm]{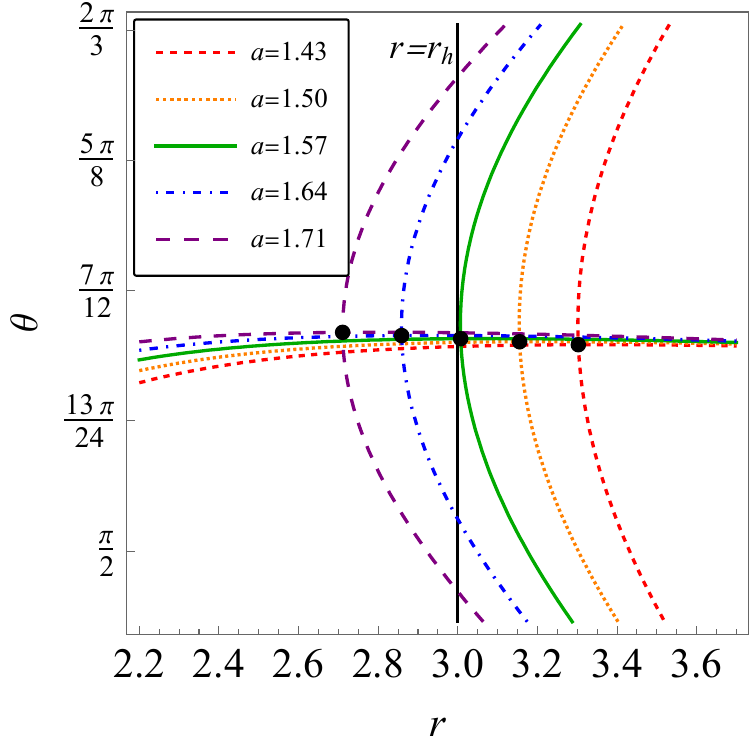}}
		\\
		\subfigure[
\label{rtheta1-}]{\includegraphics[width=7cm]{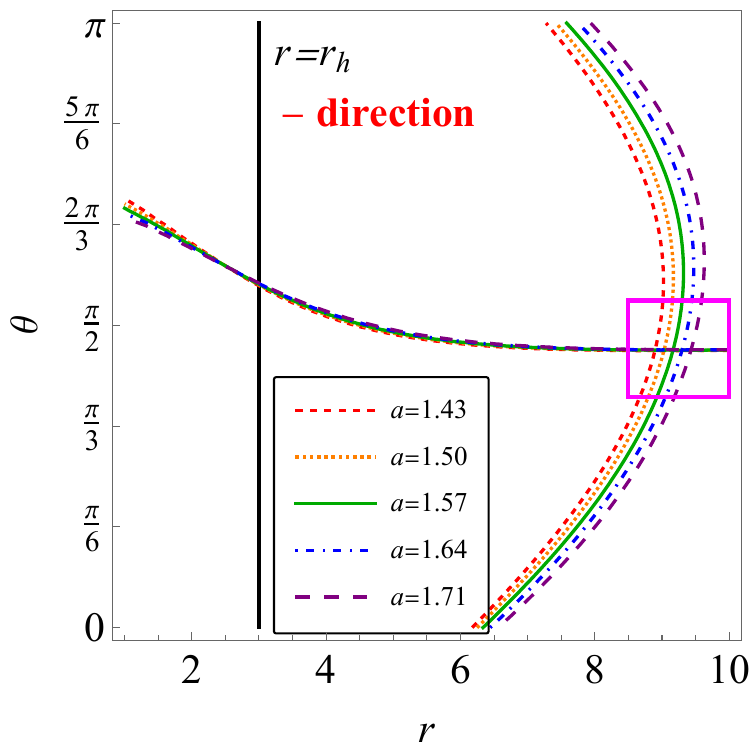}}
		\  \  \
		\subfigure[ \label{rtheta2-}]{\includegraphics[width=7cm]{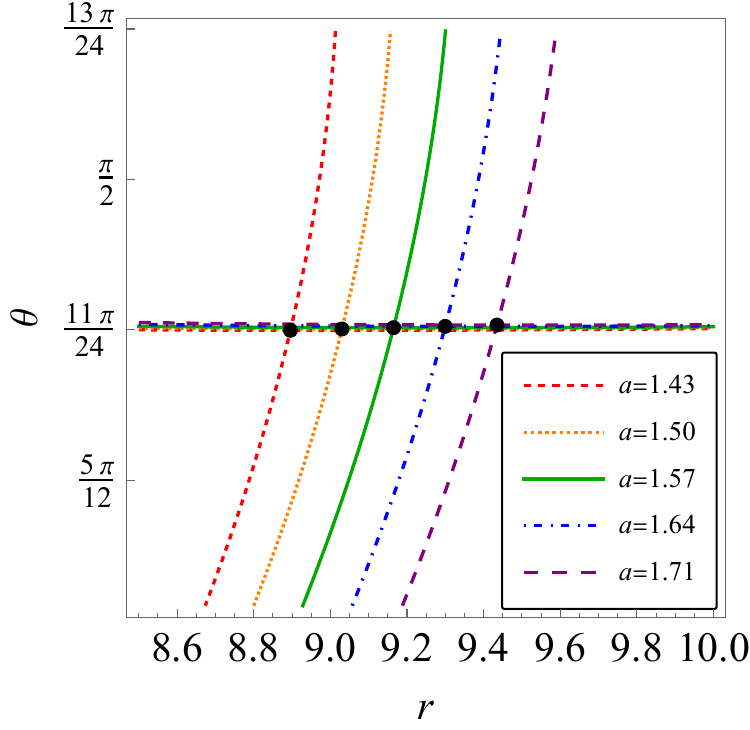}}
	\end{center}
	\caption{The schematic diagram for light ring solutions in ($r$, $\theta$) space. The intersection points marked with black dots are for the light rings. These curves are given via Eqs.~(\ref{eq_ex1LR+})-(\ref{eq_ex2LR-}). The black vertical lines denote the locations of the black hole horizons. (a) and (b) are for the prograde case, whereas (c) and (d) are retrograde case. (b) and (d) are the zooming diagrams of (a) and (c) near the light rings.}
	\label{a_r_theta}
\end{figure}

On the other hand, we focus on the location of light rings for the extremal Kerr-Newman Taub-NUT black holes. The prograde light rings are determined by
\begin{gather}
	a^4\cos ^2\theta +2a^2n\cos \theta +2\left( m-a\sin \theta \right) r-r^2+n^2=0,
	\label{eq_ex1LR+}
	\\
	-a^2\cos ^3\theta +\left( r^2+n^2+2a^2 \right) \cos \theta +2(r-m)(n+a\cos \theta )\sin \theta +2an=0.
	\label{eq_ex2LR+}
\end{gather}
For the retrograde case, the equations can be obtained by $n\rightarrow-n$ and $a\rightarrow-a$,
\begin{gather}
	a^4\cos ^2\theta -2a^2n\cos \theta +2\left( m+a\sin \theta \right) r-r^2+n^2=0,
	\label{eq_ex1LR-}
	\\
	-a^2\cos ^3\theta +\left( r^2+n^2+2a^2 \right) \cos \theta +2(r-m)(n-a\cos \theta )\sin \theta +2an=0.
	\label{eq_ex2LR-}
\end{gather}
Taking $m=3$ and $n=1$, we plot these curves determined by Eqs.~(\ref{eq_ex1LR+}) and (\ref{eq_ex2LR+}) in Fig. \ref{a_r_theta} near $a$=1.57. Then the solutions or the locations of the light rings are at the intersection points marked with black dots of these curves. For the prograde case, the results are listed in Figs. \ref{rtheta1+} and \ref{rtheta2+}. With the increase of the black hole spin, these intersection points are shifted towards to small $r$. Of particular interest is that at $a\approx1.57$, the interaction point exactly coincides with the horizon. Further increasing $a$, the point will possess smaller $r$, which indicates that the light rings for these cases will hide behind the black hole horizon. Since we only concern these light rings outside the black hole horizon, the pattern exhibits the tendency that the light rings disappear with the increase of the black hole spin. This also results in the topological phase transition near $a\approx1.57$, where the number of the light rings changes. A minor phenomenon is that, the light rings represented by these black dots slightly deviate from the equatorial plane, which is mainly caused by the breaking of the $\mathbb{Z}_2$ symmetry.

For the retrograde case, the results are also shown in Figs. \ref{rtheta1-} and \ref{rtheta2-}. It is obvious that these light rings marked with black dots are all outside the black hole horizon near $a=1.57$. This also holds for other allowed black hole spin. Therefore, we always observe one unstable light ring, indicating topological number $W=-1$ and no topological phase transition occurs.

Note that we here discuss the case for varying black hole spin, while the case considered for varying the NUT charge can be found in Appendix \ref{Example3}.

\section{Conclusions and discussions}\label{conclusion}

In this paper, we considered the topology of light rings in the backgrounds of the Kerr-Newman Taub-NUT black holes. Our study extends previous work to a black hole without the $\mathbb{Z}_2$ symmetry. The corresponding light rings are found to deviate from the equatorial plane with non-vanished NUT charge $n$. Moreover, the topology is also examined for the extremal spinning black holes.

Our first concern focuses on the non-extremal Kerr-Newman Taub-NUT black holes. By considering the direction of vector on the four segments, at $r=r_{h}$ and $\infty$, $\theta=0$ and $\pi$, of the boundary, we observed the topological number $W=-1$ for both the prograde and retrograde light ring cases. This implies that at least one unstable light ring exists. Such result is the same as that of the non-extremal Kerr-like black hole with $\mathbb{Z}_2$ symmetry. Therefore, the $\mathbb{Z}_2$ symmetry has no influence on the number of the light rings, while it indeed affects their locations. Furthermore, we also confirmed that there is one and only one light ring for the non-extremal Kerr-Newman Taub-NUT black holes.

In order to uncover the detailed influence of the $\mathbb{Z}_2$ symmetry on the light ring topology. We further divided the considered whole parameter space $X$ in $\theta-r$ plane into the upper half-plane $X_{u}$ and lower half-plane $X_{d}$. Accordingly, the corresponding boundary is separated to $C=C^{u}\cup C^{d}$. Further considering the direction of the vector along these segments of the boundary, we found that, for the prograde light ring, $W_d=0$ and $W_u=-1$ for positive NUT charge, while $W_d=-1$ and $W_u=0$ for negative NUT charge. This also uncovers that the positive or negative NUT charge shifts the light ring above or below the equatorial plane. On the other hand, for the retrograde light ring, the results reverse. Nevertheless, the total topological number $W=W_d+W_u$ keeps unchanged.

Moreover, we also performed our topology study to the extremal Kerr-Newman Taub-NUT black holes. Although it is known that there must exist one unstable light ring for the retrograde case, whether it still holds for the prograde case is still unclear. Here we expanded the vector near the black hole degenerate horizon. Further combining with the analysis, we clearly showed that the topological pattern is critically dependent of the leading term of $v_{r}^\pm$ at the zero point of $v_{\theta}^\pm$ on the black hole horizon. Different signs of the leading term gives different topological number. Meanwhile, the zero point of the first-order term represents a phase transition.

Based on these results, we took the extremal Kerr-Newman black hole as an first example. For the prograde case, the topological number $W=-1$ or 0 for a slowly spinning ($0<a<m/2$) or rapidly spinning ($a\geq m/2$) extremal black hole, which indicates that the light ring only exists for the slowly spinning black holes. And a topological phase transition occurs at $a=m/2$. However, for the retrograde case, the unstable light ring always exists. When we applied this approach to the extremal Kerr-Newman Taub-NUT black holes, we observed the similar results. The retrograde light ring exists for an arbitrary allowed spin, while the prograde light ring encounters a topological phase transition at a certain black hole spin. One significant difference is that the locations of these light rings are off the equatorial plane.

As a brief summary, we found several characterized topological properties of the light rings in this paper. First, the $\mathbb{Z}_2$ symmetry affects the location of the light ring, while has no contribute to the topological number. Second, the topological properties are similar for both the prograde and retrograde light rings in non-extremal spinning black hole backgrounds. Third, there is potential topological phase transition for the prograde light ring in the extremal spinning black hole backgrounds. While the retrograde light ring always exists. Although we mainly dealt with the Kerr-Newman Taub-NUT black holes, our study still uncovers some universal topological properties for the light rings even without  $\mathbb{Z}_2$ symmetry. These shall have enlightening significance on understanding the light rings for other spinning black holes.

\section*{Acknowledgments}
This work was supported by the National Natural Science Foundation of China (Grants No. 12075103, No. 12247101).

\appendix
\section{The light rings with $g_{\phi \phi}=0$ }\label{the light rings with gphiphi=0}
In this appendix, we would like to discuss the case of $g_{\phi \phi} =0$ appearing in Eqs. (\ref{eq_V}) or (\ref{eq_Vexpression}). After a simple calculation, we will show that $g_{\phi \phi} =0$ is contained in Eq.~(\ref{eq_cond}). We begin with Eq. (\ref{eq_Vexpression})
\begin{equation}
	V=-\frac{\Phi ^2}{D}g_{\phi \phi}(E/\Phi -H_+)(E/\Phi -H_-).
	\label{eq_VexpressionAppendix}
\end{equation}
Employing Kerr-Newman Taub-NUT black hole metric (\ref{eq_ds^2}), we can easily get
\begin{gather}
	g_{\phi \phi}=\frac{\varLambda _+\left( r,\theta \right) \varLambda _-\left( r,\theta \right)}{r^2+\left( n+a\cos \theta \right) ^2},\\
	H_\pm=\frac{a\sin \theta \pm \sqrt{\Delta}}{\varLambda _\pm \left( r,\theta \right)},
\end{gather}
where
\begin{equation}
	\varLambda _\pm \left( r,\theta \right) =\left( r^2+n^2+a^2 \right) \sin \theta \pm \left( a\sin ^2\theta -2n\cos \theta \right) \sqrt{\Delta} .
\end{equation}
It is easy to find that $g_{\phi \phi} = 0$ corresponds to
\begin{equation}
	\varLambda _+(r,\theta)  = 0 \quad \text{or} \quad \varLambda _-(r,\theta)  = 0.
\end{equation}
Their roots of ($r$, $\theta$) are different if they exist. Assuming $(r, \theta) = (\hat{r},\hat{\theta})$ is a solution of $\varLambda _+(r,\theta)  = 0$, the Eq. (\ref{eq_VexpressionAppendix}) can be simplified as
\begin{equation}
	\left. V \right|_{r=\hat{r}, \theta =\hat{\theta}}=2\frac{\Phi ^2}{D}\frac{\left( \hat{r}^2+n^2+a^2 \right) \sin \hat{\theta}}{\hat{r}^2+( n+a\cos \hat{\theta} ) ^2}\left( a\sin \hat{\theta} +\sqrt{\Delta} \right) \left.(E/\Phi -H_-)\right|_{r=\hat{r}, \theta =\hat{\theta}}.
	\label{eq_VE/Phi-H_-}
\end{equation}
Moreover, due to $\varLambda _+(\hat{r},\hat{\theta})  = 0$, we have
\begin{equation}
	\sqrt{\Delta}=\frac{\left( \hat{r}^2+n^2+a^2 \right) \sin \hat{\theta}}{2n\cos \hat{\theta} -a\sin ^2\hat{\theta}}.
\end{equation}
Thus we get
\begin{equation}
	a\sin \hat{\theta} +\sqrt{\Delta}=\frac{\hat{r}^2+(n+a\cos \hat{\theta})^2 }{2n\cos \hat{\theta} -a\sin ^2\hat{\theta}}\sin \hat{\theta} >0,
\end{equation}
which implies that the coefficient of Eq. (\ref{eq_VE/Phi-H_-}) is greater than zero. Therefore, the conditions of the light ring $V = 0$ and $\partial_\mu V = 0$ indicate that $\varLambda _+(r,\theta)  = 0$ corresponds to ``-" light ring, i.e,
\begin{equation}
	E/\Phi = H_-  \quad \text{and} \quad \partial_\mu H_- =0, \quad \mu = r, \theta .
\end{equation}
Furthermore, $\partial_{\mu}^2V > 0 $ (stable) and $\partial_{\mu}^2V < 0 $ (unstable) for light ring correspond to $\partial_{\mu}^2H_- < 0 $ and $\partial_{\mu}^2H_- > 0 $, respectively. These show that $\varLambda _+(r,\theta)  = 0$ can be determined by the retrograde case. Similarly, $\varLambda _-(r,\theta)  = 0$ can also be determined by the prograde case. As a result, one can conclude that the case $g_{\phi \phi} = 0$ for light ring is contained in Eq.~(\ref{eq_cond}).

\section{Vector and topological number for $n\neq0$}\label{Figure of V F}

We aim to show the pattern of vector, and the corresponding topological number for certain NUT charge $n$.

\begin{figure}
	\begin{center}
		\subfigure[ \ $(m,q,a,n) = (3,1,1,0)$ \label{TNB1}]{\includegraphics[width=7cm]{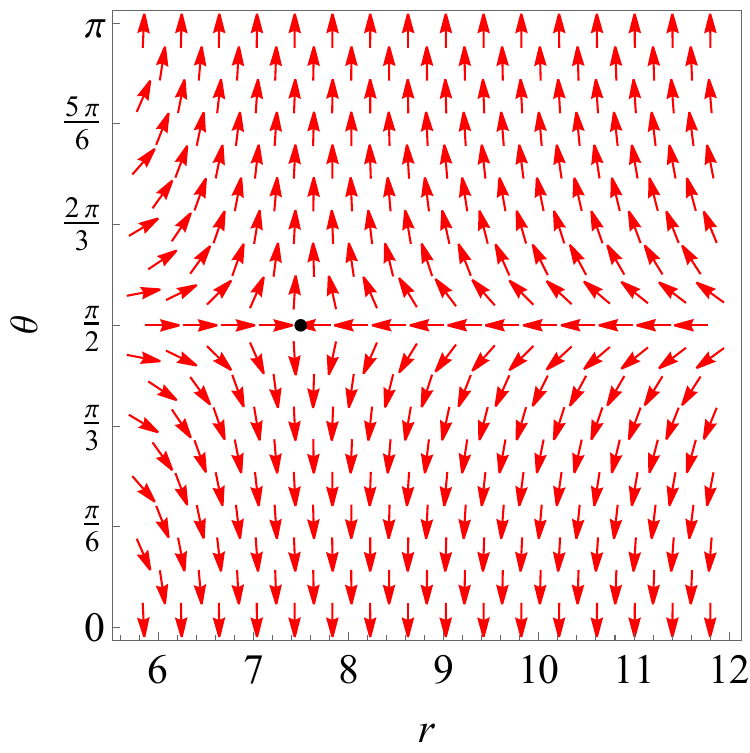}}
		\  \
		\subfigure[\ $(m,q,a,n) = (3,1,1,1)$
		 \label{TNB2}]{\includegraphics[width=7cm]{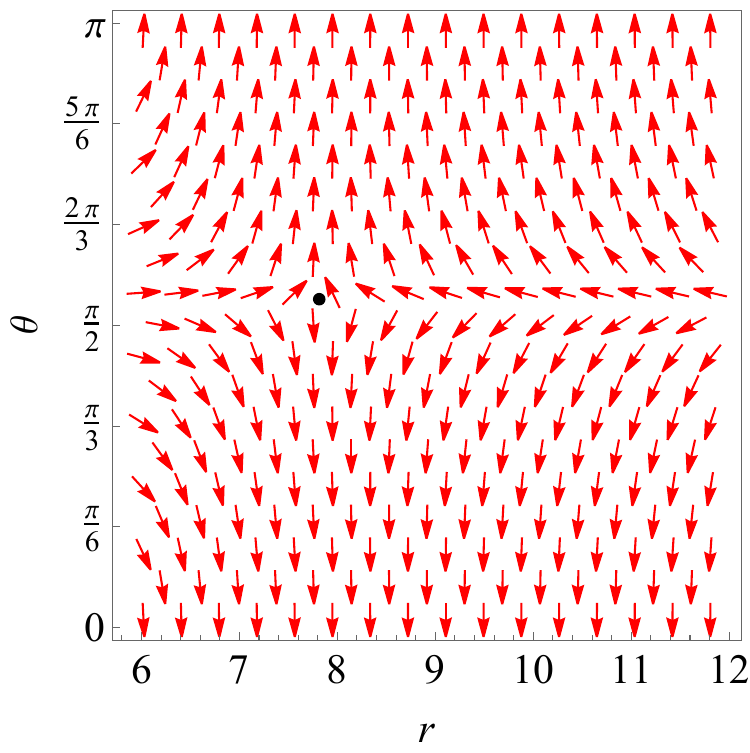}}
		\  \
		\subfigure[\ $(m,q,a,n) = (3,1,1,2)$ \label{TNB3}]{\includegraphics[width=7cm]{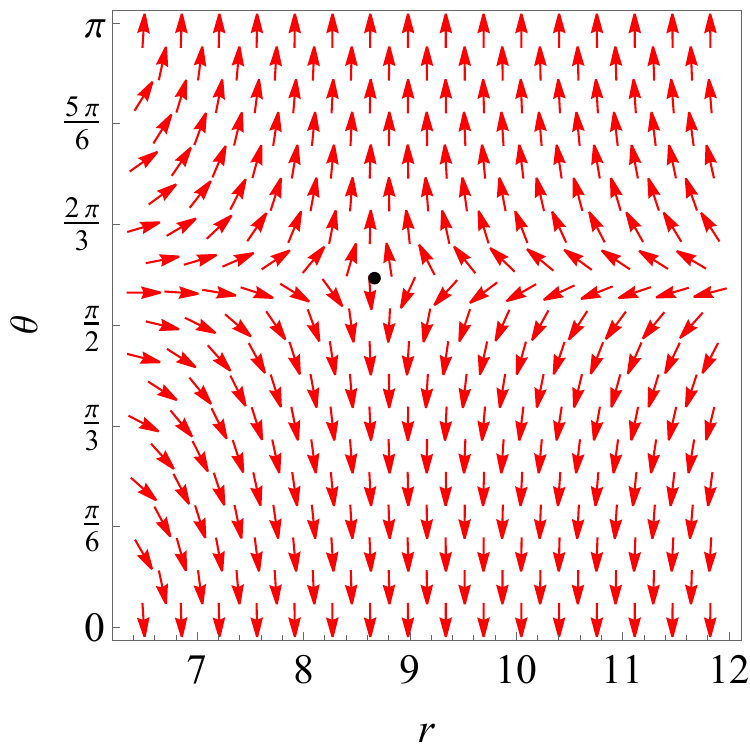}}
		\  \
		\subfigure[\ $(m,q,a,n) = (3,1,1,-1)$
		 \label{TNB4}]{\includegraphics[width=7cm]{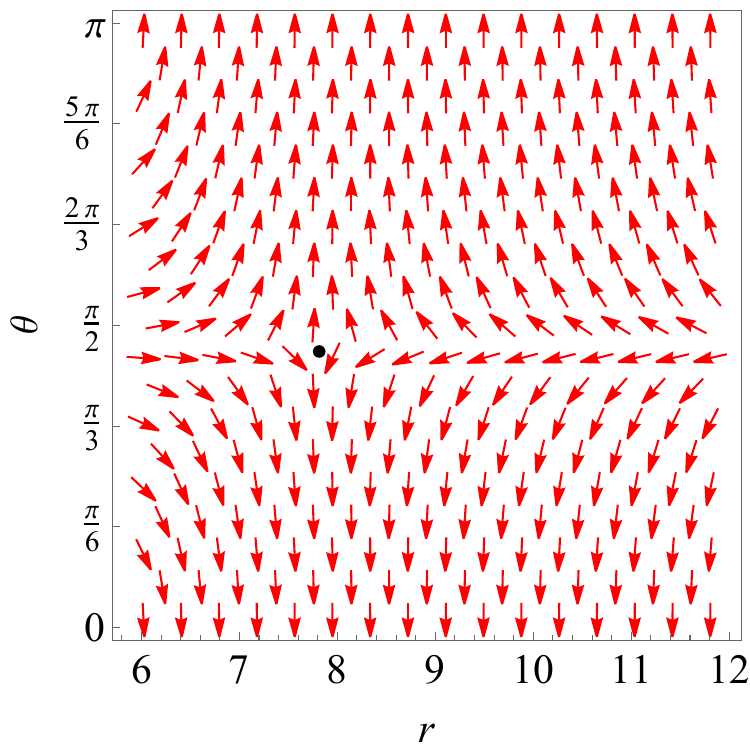}}
	\end{center}
	\caption{The red arrows represent $v$ (normalized to unity) on a portion of the ($r$, $\theta$) plane with $m=3, q=1, a =1$ for the prograde case. Black dots denote the zero points of $v$, which correspond to the prograde light rings. (a) $n=0$. (b) $n=1$. (c) $n=2$. (d) $n=-1$.}
	\label{TNB}
\end{figure}

The behaviors of the vector (normalized to unity) are shown in Fig. \ref{TNB} for the prograde light ring with fixing $m = 3$, $a=q=1$, but varying $n$ = 0, 1, 2, and -1, respectively. From the figures, we can clearly see that the vector is outwards at $\theta=0$ and $\pi$. In particular, for the NUT charge $n$=0, the zero points marked with black dot is on the equatorial plane with $\theta=\pi/2$, see Fig. \ref{TNB1}. Moreover, the zero points are shifted upwards for positive $n$ and downwards for negative $n$.

\begin{figure}
	\begin{center}
		\subfigure[ \ $(m,q,a,n) = (3,1,2,1)$ \label{cp11}]{\includegraphics[width=7cm]{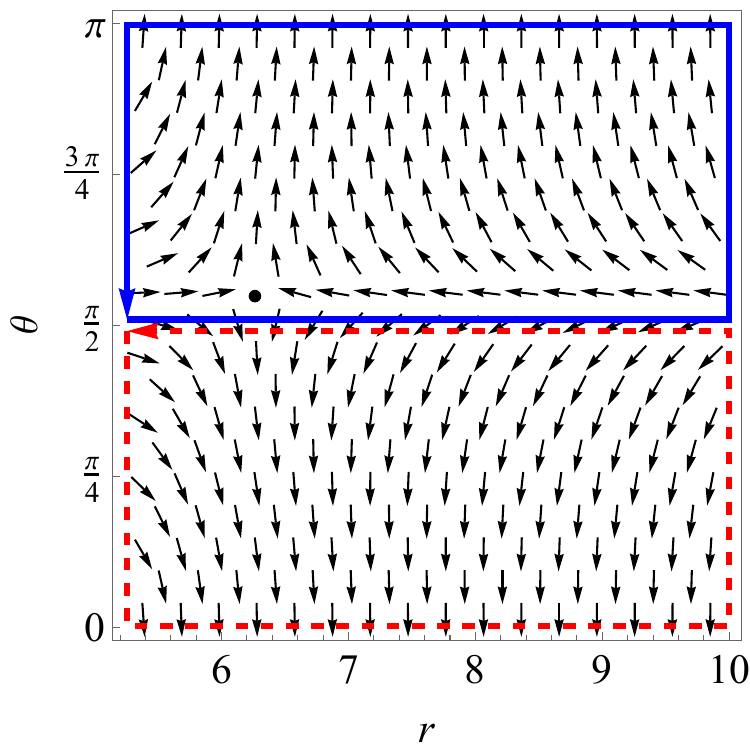}}
		\  \
		\subfigure[\ $(m,q,a,n) = (3,1,2,-1)$ \label{cp21}]{\includegraphics[width=7cm]{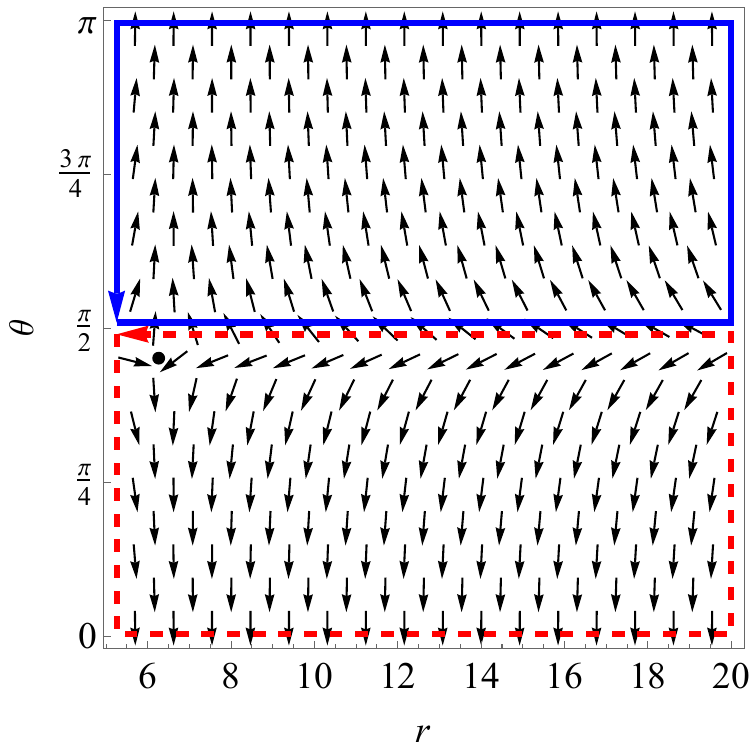}}
		\ \
		\subfigure[\ $(m,q,a,n) = (3,1,2,1)$
		 \label{cp12}]{\includegraphics[width=7cm]{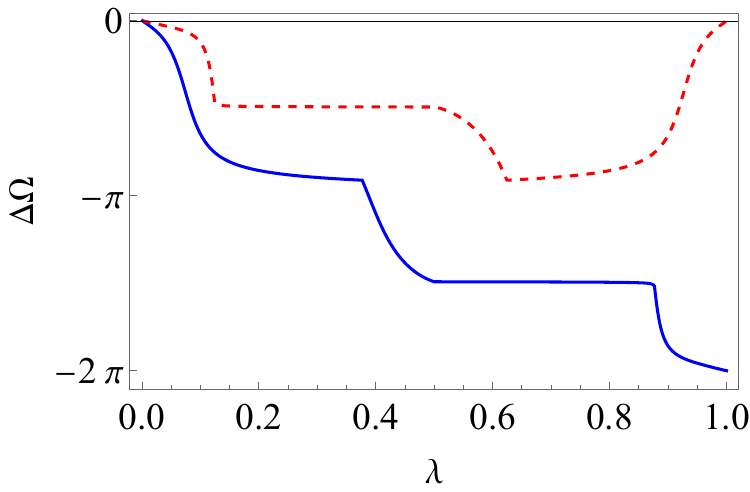}}
		\  \
		\subfigure[\ $(m,q,a,n) = (3,1,2,-1)$
		 \label{cp22}]{\includegraphics[width=7cm]{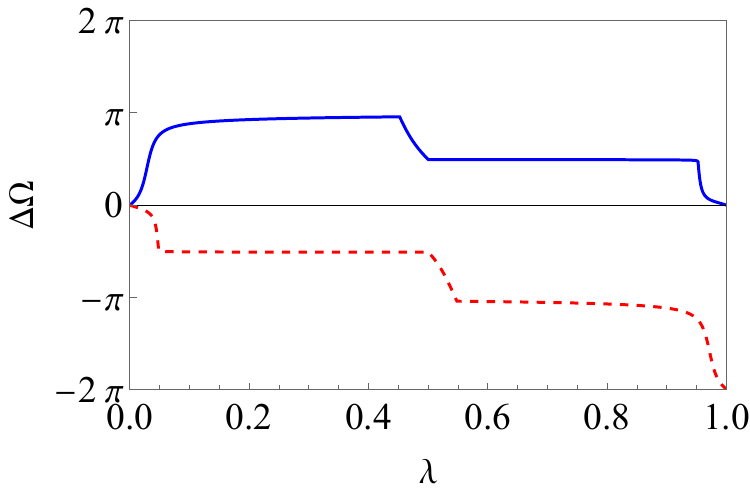}}
	\end{center}
\caption{Vector $v$ (normalized to unity) on a portion of the ($r$, $\theta$) plane and the deflection angle $\Delta\Omega$ along the boundary. We set $m$=3, $q$=1, and $a=2$. (a) $n$=1. (b) $n$=-1. (c) $n$=1. (d) $n$=-1. Black dots denote the zero points of $v$, which correspond to the prograde light rings. $\lambda$ is a length parameter of the boundaries shown in (a) and (b). The winding number is given by $w=\Delta\Omega/2\pi$. Note that there exists no zero points for large $r$.}
	\label{cp}
\end{figure}

For positive and negative $n$, we exhibit the the vector (normalized to unity), and the contours $C^u$ and $C^d$ in Figs. \ref{cp11} and \ref{cp21} represented by the blue solid and red dashed lines. We also calculate the deflection angle $\Delta\Omega$ as the function of $\lambda$, and the results are shown in Figs. \ref{cp12} and \ref{cp22}. If $\lambda$ varies from 0 to 1, one shall make one loop counterclockwise along the contour. Note that the point relating $\lambda=0$ is at $(r_h, \pi/2)$. The topological number shall be obtained via $W=\Delta\Omega(\lambda=1)/2\pi$. Therefore, it is clear that $W^{u}=-1$ and $W^{d}=$0 for positive $n$, and $W^{u}=$0 and $W^{d}=-1$ for negative one. For the retrograde case, the results will reverse. Interestingly, $W=W^{u}+W^{d}$ is also obvious. Therefore, we confirm the results given in Table \ref{T1} by taking the specific examples.

\section{Proof for inequality $\partial_\theta g$($r$, $\theta$)$ < 0$ at $g $($r$, $\theta$)=0}
\label{proof2}

For convenience, we introduce $x = \cos \theta$. Then the Eq.~(\ref{eq_g}) becomes
\begin{equation}
	g(r,x) = -a^2 x^3  +(r^2+n^2+2a^2) x + 2 a n + 2(n+a x) \sqrt{1-x^2} \sqrt{\Delta},
	\label{eq_gx}
\end{equation}
where $x \in (-1,1)$. To simplify the discussion, we denote $ g (x) \equiv g (r, x)$. Then the inequality $\partial_\theta g(r,\theta) < 0$ at $g (x)=0$ turns to
\begin{equation}
	\partial_{x} g(x) > 0.
	\label{pro}
\end{equation}
The case of $n = 0$ corresponds to the Kerr-Newman black hole whose uniqueness of the light ring (for both the prograde and retrograde cases) solution is obvious. Therefore, we mainly focus on the case of $n \ne 0$, and for simplicity, we consider the following two cases: $a=0$ and $a>0$.

\subsection{Case: $a = 0$}\label{pro1}
When $a$ vanishes, $g(x)$ will be simplified as
\begin{equation}
	g\left( x \right) =x\left( n^2+r^2 \right) +2n\sqrt{1-x^2}\sqrt{\Delta}.
\end{equation}
At the boundaries $x=-1$ and $x=1$, we respectively have
\begin{eqnarray}
	g\left(x=-1 \right) &=&-\left( r^2+n^2 \right) , \label{ado}\\
 	 g\left(x=1 \right) &=&r^2+n^2.
	\label{b0}
\end{eqnarray}
Obviously, $g\left(x=-1 \right)$ and $g\left(x=1 \right) $ are negative and positive, respectively. The first and second derivatives of $g(x)$ are given by
\begin{align}
	\partial _xg\left( x \right) &=n^2+r^2-\frac{2nx\sqrt{\Delta}}{\sqrt{1-x^2}},
	\\
	\partial _{x}^{2}g\left( x \right) &=-\frac{2n\sqrt{\Delta}}{\left( 1-x^2 \right) ^{3/2}}.
\end{align}
For further discussion, we divide the NUT charge into the following two cases: $n>0$ and $n<0$.
\begin{itemize}
	\item $n>0$. In this case, $\partial_x^2 g(x) < 0$ is obvious, which indicates that $\partial_x g(x)$ monotonically decreasing with $x$. On the other hand, we can find $\left. \partial_x g(x) \right|_{x \rightarrow -1} > 0$ and $\left. \partial_x g(x) \right|_{x \rightarrow 1} < 0$. Thus, $g(x)$ increases at first and then decreases with $x$. Further considering Eqs.~(\ref{ado}) and (\ref{b0}), $g(x)=0$ has only one root and $\partial _xg\left( x \right)$ must be positive at the zero point of $g(x)=0$.
	\item $n<0$. In this case, $\partial_x^2 g(x) > 0$ is obvious, which indicates that $\partial_x g(x)$ monotonically increasing with $x$. On the other hand, we can find $\left. \partial_x g(x) \right|_{x \rightarrow -1} < 0$ and $\left. \partial_x g(x) \right|_{x \rightarrow 1} > 0$. Thus, $g(x)$ decreases at first and then increases with $x$. Further combining with Eqs.~(\ref{ado}) and (\ref{b0}), $g(x)=0$ has only one root and $\partial _xg\left( x \right)$ must be positive at the zero point of $g(x)=0$.
\end{itemize}
Therefore, one can obtain the inequality (\ref{pro}) for $a=0$.

\subsection{Case: $a > 0$}\label{pro2}

For positive black hole spin, we can re-scale the function $g(x)$ with $a$. In order to preserve its expression, we need to set $r/a\rightarrow r$, $n/a\rightarrow n$, $q/a\rightarrow q$, and $m/a\rightarrow m$. Or equivalently, we can set $a=1$. Then the Eq.~(\ref{eq_gx}) shall be
\begin{equation}
		g(x) = - x^3  +(r^2+n^2+2) x + 2  n + 2(n+x) \sqrt{1-x^2} \sqrt{\Delta}.
\end{equation}
From this formula, we can obtain
\begin{align}
	g( -1 ) &=-\left( r^2+\left( n-1 \right) ^2 \right),
	\\
	g( 1 ) &=r^2+\left( n+1 \right) ^2,
	\\
	g( 0 ) &=2n\left( 1+\sqrt{\Delta} \right).
\end{align}
The corresponding derivatives of $g(x)$ are
\begin{align}
	\partial _x g(x ) &=-3x^2+r^2+n^2+2-\frac{2\left( 2x^2+nx-1 \right) \sqrt{\Delta}}{\sqrt{1-x^2}},
\\
	\partial _{x}^{2} g( x ) &=-6x-\frac{2\left( -2x^3+3x+n \right) \sqrt{\Delta}}{\left( 1-x^2 \right) ^{3/2}},
\\
	\partial _{x}^{3}g( x ) &=-6-\frac{6(nx+1)\sqrt{\Delta}}{\left( 1-x^2 \right) ^{5/2}}.	
\end{align}
For further discussion, we divide the NUT charge into the following three cases: $n>1$, $0<n\leq1$, and $n<0$.
The discussion is as follows.
\begin{itemize}
	\item $0<n\leq 1$. We have $\partial _{x}^{3}g( x )<$ 0, $\left. \partial_x^2 g(x) \right|_{x \rightarrow -1} > 0 $, and $\left. \partial_x^2 g(x) \right|_{x \rightarrow 1} < 0 $, which implies that $\partial _{x}^{2}g\left( x \right)$ has one zero point and $\partial_x g(x)$ first increases and then decreases. Furthermore, we also have $\left. \partial_xg(x) \right|_{x \rightarrow 0} > 0 $, $\left. \partial_xg(x) \right|_{x \rightarrow -1} \leqslant 0 $ and $\left. \partial_xg(x) \right|_{x \rightarrow 1} < 0 $, which indicate that $g(x)$ decreases first\footnote{If $\left. \partial_xg(x) \right|_{x \rightarrow -1} = 0$ holds, the decreasing behavior will no longer exist.}, then increases, and finally decreases as $x$ varies from $-1$ to $1$. Finally, combining with $g(-1)<0$ and $g(1)>0$, we get $\partial_x g(x) > 0$ when $g(x)=0$.
	\item $n>1$. When $-1 < x \leqslant 0$, one has $\partial_x g(x) > 0$, while when $0<x<1$, $ \partial_x ^2 g(x) < 0 $. We also have $\left. \partial_x g(x) \right|  _{x \rightarrow 0} >0$ and $\left. \partial_x g(x) \right|  _{x \rightarrow 1} <0$. These mean that $g(x) $ will increase first and then decrease as $x$ varies from $- 1$ to $1$. By making use the condition $g(-1)<0$ and $g(1)>0$, we can obtain $\partial_x g(x) > 0$ when $g(x)=0$.
	\item $n<0$. As shown above, we have confirmed $\partial_x g(x) > 0$ at $g(x)=0$ for $n>0$. The case with $n<0$ is straightforward by using the relation $g(x,n) = -g(-x,-n)$. For example, for negative $n$, we have $\partial_{-x} g(-x,-n)>0$ from above result. Further considering $\partial_{-x} g(-x,-n)=-\partial_{-x} g(x,n)=\partial_{x} g(x,n)$, one easily has $\partial_x g(x) > 0$ as expected.
\end{itemize}
In summary, we have checked $\partial_x g(x) > 0$ at $g(x)=0$ for the non-negative black hole spin.

\section{Topological phase transition with varying $n$}
\label{Example3}

Here, we would like to consider the topological number and phase transition for extremal black hole with varying the NUT charge $n$. For simplicity, we set $m=2$, $a=2$, and $\left|q\right|_{max}=10$. Due to the constraint $m^2+n^2=a^2+q^2$, we need vary $n$ from -10 to 10 as expected.

\begin{figure}
	\begin{center}
		 \label{n_Vc}{\includegraphics[width=7cm]{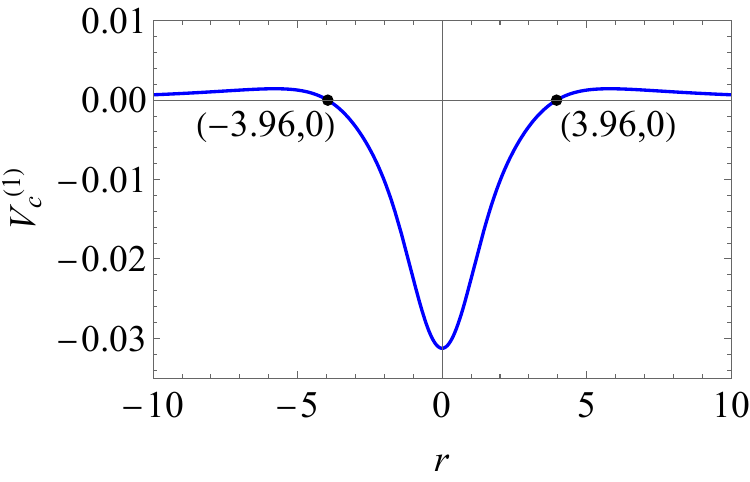}}
	\end{center}
	\caption{Quantities $V_c^{+(1)}$ (\ref{eq_Vc1}) as a function of $n$ with $m=2$ and $a =2$.}
	\label{FIG.n_Vc1}
\end{figure}

The quantity $V_c^{+(1)}$ is plotted in Fig. \ref{FIG.n_Vc1}, but $V_c^{-(1)}$ is not plotted for the reason that it is always less than zero and no phase transition exists. From this figure, we can find that there will be a phase transition when $|n|$ is around $3.96$. A simple calculation shows that, the topological number is -1 for $ \left|n\right| > 3.96$, and zero for  $\left|n\right| \leq 3.96$.

\begin{figure}
	\begin{center}
		\subfigure[
\label{rtheta1n}]{\includegraphics[width=7cm]{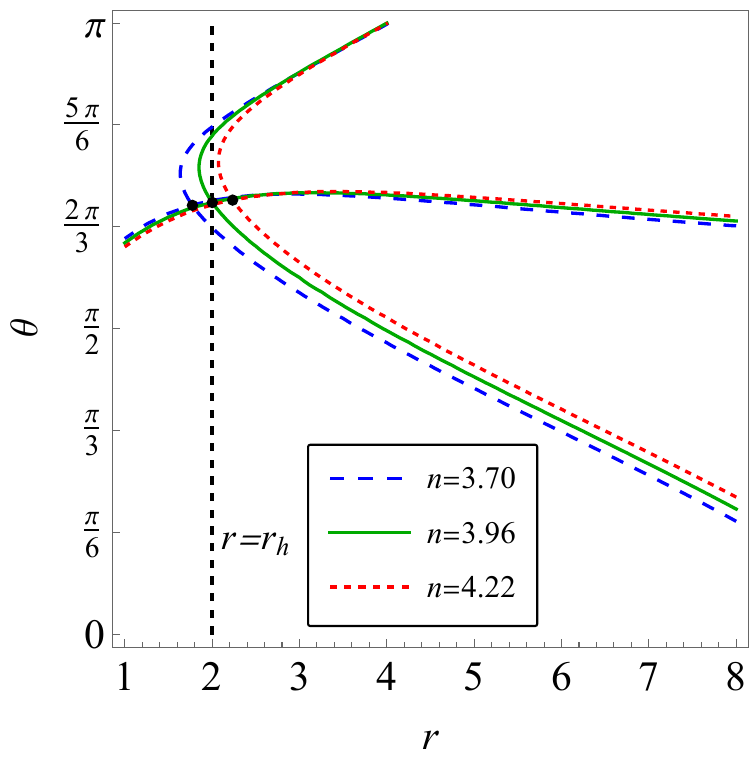}}
		\  \  \
		\subfigure[ \label{rtheta2n}]{\includegraphics[width=7cm]{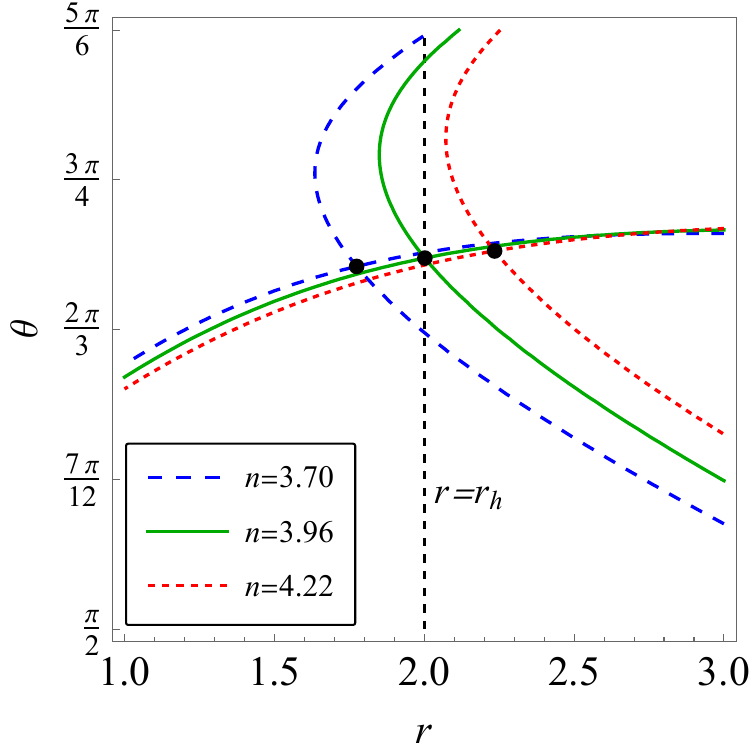}}
				\subfigure[
		 \label{rtheta3n}]{\includegraphics[width=7cm]{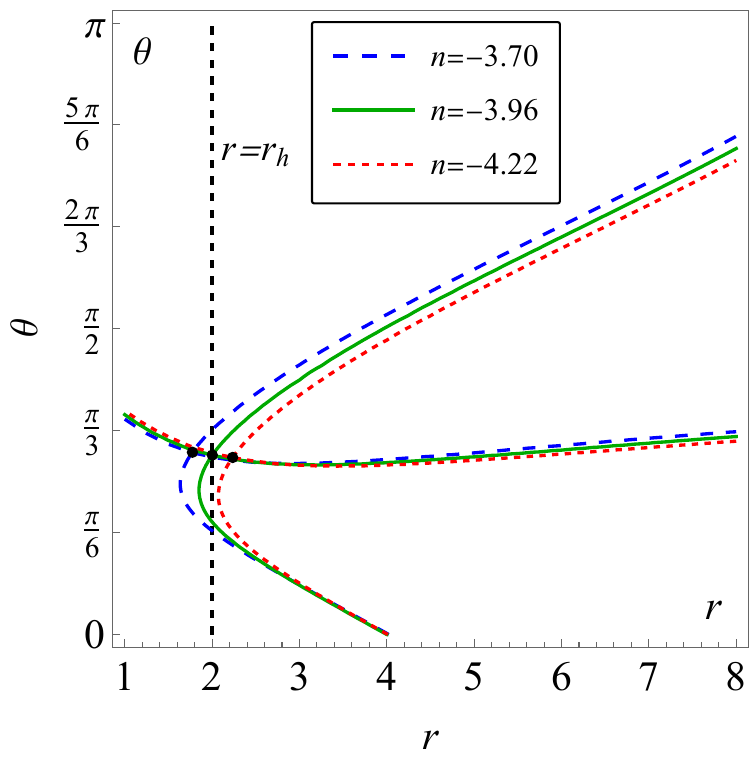}}
		\  \  \
		\subfigure[ \label{rtheta4n}]{\includegraphics[width=7cm]{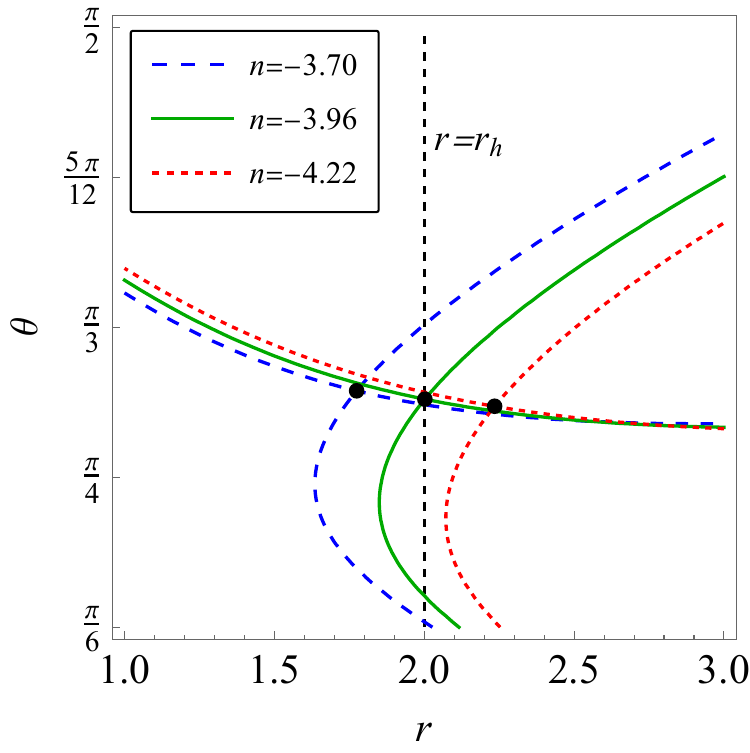}}
	\end{center}
		\caption{The schematic diagram of light ring solutions in ($r$, $\theta$) plane for the prograde light ring. The intersection points of these curves marked with black dots are for the prograde light rings. These curves are given by Eqs.~(\ref{eq_ex1LR+}) and (\ref{eq_ex2LR+}). The black vertical lines represent the locations of the black hole horizons.}
	\label{n_r_theta}
\end{figure}

On the other hand, the prograde light ring can be obtained by the intersections of these two kinds curves determined by Eqs.~(\ref{eq_ex1LR+}) and (\ref{eq_ex2LR+}). Taking $n=$3.70, 3.96, 4.22 and $n =$-3.70,  -3.96, 4.22, we show these curves in Fig. \ref{n_r_theta}. The light rings are marked with the black dots. It is clear that when $\left|n\right| > 3.96$, there are light rings outside the black hole horizon. While when $\left|n\right| < 3.96$, the light rings are behind the event horizon, leading to the disappearance of these light rings. This clearly exhibits that there is a topological phase transition near $n\approx3.96$ characterizing whether the light ring exists or not.

\end{document}